\documentclass[showpacs,twocolumn,eqsecnum,prb,amsmath,amssymb]{revtex4}
\usepackage{graphicx}% Include figure files
\usepackage{dcolumn}% Align table columns on decimal point
\usepackage{bm}% bold math
\begin{document}
\title{
Application of Kondo-lattice theory to
the Mott-Hubbard metal-insulator crossover\\ 
in disordered cuprate oxide superconductors 
}
\author{Fusayoshi J. Ohkawa}
\affiliation{Division of Physics, Graduate School of Science, 
Hokkaido University, Sapporo 060-0810, Japan}
\email{fohkawa@phys.sci.hokudai.ac.jp}
%\date{\today}
\received{6 May 2004}
\begin{abstract} 
A theory of Kondo lattices is applied to the crossover between
local-moment magnetism and itinerant-electron magnetism in the $t$-$J$
model on a quasi-two dimensional lattice. The Kondo temperature
$T_K$ is defined as a characteristic temperature or energy scale
of local quantum spin fluctuations.  Magnetism
with $T_N \gg T_K$, where $T_N$ is the N\'{e}el temperature, is
characterized as local-moment one, while magnetism with  $T_N\ll T_K$
is characterized as itinerant-electron one. 
The Kondo temperature, which also gives a measure of the strength of
the quenching of magnetic moments, is renormalized by the Fock term of
the superexchange interaction. Because the renormalization depends on
life-time widths $\gamma$ of quasiparticles  in such a way that  $T_K$
is higher for smaller 
$\gamma$, $T_N$ can be controlled by disorder. The asymmetry of
$T_N$ between electron-doped and hole-doped cuprates must mainly arise
from that of disorder; an almost symmetric behavior of
$T_N$ must be restored if we can  prepare hole-doped and
electron-doped cuprates with similar degree of disorder to each other.
Because effective disorder is enhanced by magnetic fields in Kondo
lattices, antiferromagnetic ordering must be induced by magnetic fields
in cuprates that exhibit large magnetoresistance.

\end{abstract}
%%%%%%
% 71.10.-w Theories and models of many electron systems
% 71.27.+a Strongly correlated electron systems; heavy fermions
% 71.30.+h Metal-insulator transitions 
%          and other electronic transitions
% 74.20.-z Theories and models of superconducting state
% 75.10.-b General theory and models of magnetic ordering
% 75.10.Lp Band and itinerant models
% 75.30.Et Exchange and superexchange interactions
% 75.30.Kz Magnetic phase boundaries
% 75.30.Gw Magnetic anisotropy
% 75.50.-y Studies of specific magnetic materials
%%%%%%%%%%%%%%%%%%%%%%%%%%%%%%%%%%%%%%%%%%%%%%%%%%%%%%%
\maketitle

\section{Introduction}\label{SecIntroduction}

The discovery\cite{bednorz} in 1986 of high transition-temperature
(high-$T_c$) superconductivity in  cuprate oxides has revived intensive
and extensive studies on strong electron correlations because it
occurs in the vicinity of the  Mott-Hubbard metal-insulator transition
or crossover. Cuprates with no dopings are Mott-Hubbard insulators,
which exhibit antiferromagnetism at low temperatures. When electrons
or {\it holes} are doped, they show the metal-insulator crossover.
However, the crossover is asymmetric between electron-doped and
hole-doped cuprates;\cite{asymmetric} the insulating and
antiferromagnetic phase is much wider as a function of dopings in
electron-doped cuprates than it is in hole-doped cuprates.  Because
superconductivity appears in a metallic phase adjacent to the
insulating phase, clarifying what causes the asymmetry  is one of the
most important issues to settle the mechanism of high-$T_c$
superconductivity itself among various proposals.

In 1963, three distinguished theories on electron correlations in a
single-band model, which is now called the Hubbard model, were
published by Kanamori,\cite{Kanamori} Hubbard,\cite{Hubbard} and
Gutzwiller.\cite{Gutzwiller} Two theories among them are directly
related with the transition or crossover.  According to Hubbard's
theory,\cite{Hubbard} the band splits into two subbands called the
lower and upper Hubbard bands.  According to Gutzwiller's
theory,\cite{Gutzwiller} with the help of the Fermi-liquid
theory,\cite{Luttinger1,Luttinger2} a narrow quasiparticle band 
appears on the chemical potential;  we call it Gutzwiller's band in
this paper.  When we take both of them, we can argue that 
the density of states must be of a three-peak structure, Gutzwiller's
band on the chemical potential between the lower and upper Hubbard
bands.  This speculation was confirmed in a previous
paper.\cite{OhkawaSlave}  The Mott-Hubbard splitting occurs in both
metallic and insulating phases as long as the onsite repulsion $U$ is
large enough, and Gutzwiller's band is responsible for metallic
behaviors.  Then, we can argue that a metal with almost half filling
can become an insulator only when a gap opens in Gutzwiller's band or
that it can behave as an insulator  when life-time widths of
Gutzwiller's quasiparticles are  so large that they can play no
significant role.

Brinkman and Rice\cite{Brinkman} considered the transition  at $T=0$~K  and the
just half filling as a function of $U$  in Gutzwiller's
approximation.\cite{Gutzwiller} They showed that the effective mass
$m^*$ and the static homogeneous susceptibility
$\chi_s(0,{\bf q}\rightarrow 0)$ diverge at a critical $U_c$.  Their
result implies that the ground state for $U>U_c$ must be a
Mott-Hubbard  insulator and the metal-insulator transition is of second
order. In general, an order parameter appears in a second-order
transition. However, there is no evidence that any order parameter
appears  in this transition. The absence of any order parameter
contradicts the opening of gaps.  The transition is caused by the
disappearance of Gutzwiller's band; the divergence of $m^*$ is one of
its consequences.  It is interesting to examine beyond Gutzwiller's
approximation, within the Hilbert subspace restricted within
paramagnetic states, whether a hidden order parameter exists, 
whether  the critical
$U_c$ is finite or infinite, and whether the transition turns out to a
crossover. It is also interesting to examine  of which order the
transition is,  second order, first order, or crossover, at non-zero
temperatures, where itinerant electrons and holes are thermally
excited across the Mott-Hubbard gap.

It is also an interesting issue how the transition or crossover occurs
as a function of electron or dopant concentrations.  Once holes or
electrons are doped into the Mott-Hubbard insulator that is just  half
filled, it must become a metal; unless  a gap  opens in Gutzwiller's
band, there is no reason  why doped {\it hole} or electrons are
localized in a periodic system. No metal-insulator transition can
occur at nonzero concentrations of dopants  even if $U$ is infinitely
large.   For the just half filling, on the other hand, a system with
$U > U_c$ is  the  Mott-Hubbard insulator at $T=0$~K; $U_c$ may be
finite or infinite.   If $U_c$ is infinite,  the point at $U=+\infty$
and the just half filling is a singular point  in the phase-diagram
plane of $U$ and electron concentrations;  if $U_c$ is finite,  the
line on $U>U_c$ and the just half filling is a singular line. At
$T=0$~K, a  system is an insulator only at the singular point or on
the singular line  while it is a metal in the other whole region.
However, either of these phase diagrams is totally different from
observed ones. For example, cuprates with small amount of dopants are
Mott-Hubbard insulators. When enough holes or electrons are doped, the
insulators become paramagnetic metals. At low temperatures, they
exhibit antiferromagnetism in insulating phases and 
superconductivity in metallic phases. The metal-insulator crossover in
cuprates must be closely related with the disappearance of
antiferromagnetic gaps in Gutzwiller's band.  It must also be closely
related with the crossover between local-moment magnetism and
itinerant-electron magnetism.

Not only Hubbard's\cite{Hubbard} and Gutzwiller's\cite{Gutzwiller}
theories but also the previous theory\cite{OhkawaSlave} are within the
single-site approximation (SSA). Their validity 
tells that local fluctuations are responsible for the three-peak
structure.  Local fluctuations are rigorously considered in one of the
best SSA's.\cite{comSSA} Such an SSA is reduced to solving the Anderson
model,\cite{Mapping} which is one of the simplest effective
Hamiltonians for the Kondo problem. The Kondo problem has already been
solved.\cite{singlet,poorman,Wilson,Nozieres,Yamada,Yosida}   One of
the most essential physics involved in the Kondo problem is that a
magnetic moment is quenched by local quantum spin
fluctuations so that the ground states is a singlet\cite{singlet} or a
normal Fermi liquid.\cite{Wilson,Nozieres}  The Kondo temperature
$T_{K}$ is defined as a temperature or  energy scale of local quantum
spin fluctuations; it is also a measure of the strength of the
quenching of magnetic moments.  The
so called Abrikosov-Suhl or Kondo peak between two sub-peaks
corresponds to Gutzwiller's band between the lower and upper Hubbard
bands. Their peak-width or bandwidth is about $4k_BT_K$, with $k_B$ the
Boltzmann constant.

On the basis of the mapping to the Kondo problem, we  argue that a
strongly correlated electron  system on a lattice must show a
metal-insulator crossover as a function of $T$: It is a nondegenerate
Fermi liquid at $T\gg T_{K}$ because local thermal spin fluctuations
are dominant, while it is a Landau's normal Fermi liquid at $T\ll
T_{K}$ because local quantum spin fluctuations are dominant and 
magnetic moments are quenched by them. Local-moment magnetism occurs
at $T\gg T_{K}$, while itinerant-electron magnetism occurs at $T\ll
T_{K}$;  superconductivity can occur only at $T\ll T_K$, that is, in
the region of itinerant electrons. The crossover implies that the
coherence or incoherence of quasiparticles plays a crucial role in the
metal-insulator crossover.  The coherence is destroyed by not only
thermal fluctuations but also disorder. Denote the life-time width of
quasiparticles by $\gamma$. When  $k_BT$ or  $\gamma$ is larger
than Gutzwiller's  bandwidth $W^* \simeq 4k_BT_K$ such as  $k_BT\agt
W^*$ or $\gamma\agt W^*$, Gutzwiller's quasiparticles are never
well-defined and they can never play a significant role; the system
behaves as an insulator. When
$k_BT\ll W^*$ and  $\gamma\ll W^*$, on the other hand, they
are well-defined and they can play a role; the system behaves as a
metal.  Disorder can play a significant role in the Mott-Hubbard
metal-insulator transition or crossover.

A theory of Kondo lattices is formulated in such a way that an {\it
unperturbed} state is constructed in one of the best SSA's\cite{comSSA}
and intersite terms are perturbatively considered. It has already been
applied to not only  typical issues on electron correlations such as
the Curie-Weiss law of itinerant-electron magnets,\cite{ohkawaCW,miyai}
ferromagnetism induced by magnetic fields or
metamagnetism,\cite{Satoh} and itinerant-electron antiferromagnetism
and ferromagnetism,\cite{ohkawaCW,antiferromagnetism,ferromagnetism}
but also high-$T_c$ superconductivity, the mechanism of
superconductivity,\cite{OhkawaSC1,OhkawaSC2,OhkawaSC3} the opening of
pseudogaps,\cite{OhkawaPseudogap} the softening of
phonons,\cite{OhkawaEl-ph} and kinks in the quasiparticle
dispersion.\cite{OhkawaEl-ph}  Early
papers\cite{Ohkawa87SC-1,Ohkawa87SC-2,Ohkawa90SC-3} on $d\gamma$-wave
high-$T_c$ superconductivity, including the earliest two
ones\cite{Ohkawa87SC-1,Ohkawa87SC-2} published in 1987, can also be
regarded within the theoretical framework of Kondo lattices. One of
the purposes of this paper  is to apply the theory of Kondo lattices
to  the crossover between local-moment magnetism and
itinerant-electron magnetism. The other purpose is to show that the
asymmetry of the N\'{e}el temperature $T_N$  between electron-doped
and hole-doped cuprates can arise from that of disorder.  This paper
is organized as follows:  The theory of Kondo lattices is reviewed in
Sec.~\ref{SecKondo-lattice}.  Effects of the coherence of
quasiparticles on $T_N$ is studied in Sec.~\ref{SecTN}. The asymmetry
of $T_N$ in cuprates is examined in Sec.~\ref{SecDiscussion}.
Conclusion is given in Sec.~\ref{SecConclusion}. The selfenergy of
quasiparticles in disordered Kondo lattices is studied in
Appendix~\ref{SecDisorder}.  A possible mechanism for the deviation of
the Abrikosov-Gorkov theory\cite{abrikoso-gorkov} is studied in
Appendix~\ref{SecSuper}.

\section{Kondo-lattice theory}
\label{SecKondo-lattice}

\subsection{Renormalized SSA}
\label{SecRenomalizedSSA}

We consider the $t$-$J$ or $t$-$t^\prime$-$J$ model on a simple square
lattices with lattice constant $a$:\cite{exchByOhkawa}
\begin{eqnarray}\label{Eqt-Jmodel}
{\cal H} &=& 
-t \sum_{\left< ij \right>\sigma} a_{i\sigma}^\dag a_{j\sigma}
-t^\prime \sum_{\left< ij \right>^{\prime}\sigma} 
a_{i\sigma}^\dag a_{j\sigma}
\nonumber \\ && \quad
- \frac1{2} J \sum_{\left< ij \right>} 
({\bf S}_i \cdot {\bf S}_j)
+ U_{\infty} \sum_{i}n_{i\uparrow}n_{i\downarrow} , \quad 
\end{eqnarray}
with  $t$  the transfer integral between  nearest
neighbors $\left< ij\right>$, $t^\prime$ between next-nearest
neighbors $\left< ij\right>^{\prime}$,
%
%\begin{equation}
${\bf S}_i = \sum_{\alpha\beta} \frac1{2} 
\left(\sigma_x^{\alpha\beta},
\sigma_y^{\alpha\beta},
\sigma_z^{\alpha\beta} \right) 
a_{i\alpha}^\dag a_{i\beta} $,
%\end{equation}
with $\sigma_x$, $\sigma_y$, and $\sigma_z$ being the Pauli matrices,
and $n_{i\sigma}=a_{i\sigma}^\dag a_{i\sigma}$.
 Because we are interested in
cuprates, we assume that $t>0$ and the superexchange interaction is nonzero
only between nearest neighbors and is antiferromagnetic;
\begin{equation}
J/|t| = - 0.3 .
\end{equation}
Infinitely large onsite repulsion,
$U_\infty/|t|\rightarrow +\infty$, is introduced in order to exclude
doubly occupied sites.  
Effects of disorder and weak three dimensionality are
phenomenologically considered in this paper.

The $t$-$J$ model (\ref{Eqt-Jmodel}) can only treat less-than-half
fillings. When we take the hole picture, we can also treat
more-than-half fillings with the model (\ref{Eqt-Jmodel}) with the
signs of $t$ and $t^\prime$ reversed.  We consider two models: a {\it
symmetric} one with
$t^\prime=0$,  and an {\it asymmetric} one with  $t^\prime/t 
\simeq -0.3$, whose precise definition is made below. In the symmetric
model, physical properties are symmetric between less-than-half and
more-than-half fillings.

We follow the previous paper\cite{OhkawaPseudogap} to treat the infinitely
large $U_\infty$. %\cite{infinteU}
%%%%%%%%%%%%%%%%%%%%%%%%%%%%%%%%%%%%%%%%%%%%%%%%%%%%%
The single-particle selfenergy $\Sigma_\sigma(i\varepsilon_n,{\bf k})$ is
divided into  a single-site term $\tilde{\Sigma}_\sigma(i\varepsilon_n)$, an
energy-independent multisite term $\Delta \Sigma_\sigma({\bf k})$, 
and an energy-dependent multisite term
$\Delta\Sigma_\sigma(i\varepsilon_n,{\bf k})$: 
%\begin{equation}
$\Sigma_\sigma(i\varepsilon_n,{\bf k}) =
\tilde{\Sigma}_\sigma(i\varepsilon_n) +
\Delta \Sigma_\sigma({\bf k}) +
\Delta\Sigma_\sigma(i\varepsilon_n,{\bf k})$.
%\end{equation}
As is discussed below, $\Delta\Sigma_\sigma({\bf k})$ is the Fock term
due to $J$. First, we take  a renormalized SSA, which includes not only 
$\tilde{\Sigma}_\sigma(i\varepsilon_n)$ but also 
$\Delta \Sigma_\sigma({\bf k})$.  The SSA is reduced to solving a mapped
Anderson model.  The mapping condition is simple:\cite{Mapping} The
onsite repulsion of the Anderson model should be $U_\infty$, and other
parameters should be determined to satisfy
\begin{equation}\label{EqMap}
\tilde{G}_\sigma (i\varepsilon_n) \!=\!
\frac1{N} \sum_{\bf k} G_\sigma^{(0)}(i\varepsilon_n,{\bf k}) ,
\end{equation}
with $\tilde{G}_\sigma (i\varepsilon_n)$ the Green function of the Anderson
model, and
%\begin{equation}
$G_\sigma^{(0)} (i \varepsilon_{n}, {\bf k})  =
1/[ i\varepsilon_n \!\!+\! \mu \!-\! E({\bf k}) 
\!-\! \tilde{\Sigma}_\sigma(i\varepsilon_n)
\!-\! \Delta\Sigma_\sigma({\bf k}) ]$.
%\end{equation}
%
Here, $\mu$ is the chemical potential, and
\begin{equation}
E({\bf k}) = -2 t \eta_{1s}({\bf k})
-2 t^\prime \eta_{2s}({\bf k}),
\end{equation}
 with 
\begin{subequations} 
\begin{eqnarray}
\eta_{1s}({\bf k}) &=& \cos(k_x a) + \cos(k_y a) ,
\\ 
\eta_{2s}({\bf k}) &=& 2\cos(k_x a) \cos(k_y a) ,
\end{eqnarray}
\end{subequations}
is the dispersion relation of unrenormalized electrons.
The single-site term $\tilde{\Sigma}_\sigma(i\varepsilon_n)$ is given by
the selfenergy of the Anderson model. It is expanded as
\begin{equation}\label{EqExpansionSigma}
\tilde{\Sigma}_\sigma(i\varepsilon_n) =
\tilde{\Sigma}(0) \!+\! (1 \!-\! \tilde{\phi}_\gamma) i\varepsilon_n
%\nonumber \\ && \quad 
\!+\! \sum_{\sigma^\prime}(1 \!-\!
\tilde{\phi}_{\sigma\sigma^\prime}) 
\Delta\mu_{\sigma^\prime} + \cdots,
\end{equation}
with $\Delta\mu_{\sigma}$
an infinitesimally  small spin-dependent chemical potential shift. Note
that 
%\begin{equation}
$\tilde{\phi}_\gamma=\tilde{\phi}_{\sigma\sigma}$.
%\end{equation}
The Wilson ratio is defined by
%\begin{equation}
$\tilde{W}_s = \tilde{\phi}_s/\tilde{\phi}_\gamma$,
%\end{equation}
with 
%\begin{equation}
$\tilde{\phi}_s=
\tilde{\phi}_{\sigma\sigma}-\tilde{\phi}_{\sigma-\sigma}$.
%\end{equation}
For almost half fillings, charge fluctuations are suppressed so that
%\begin{equation}
$\tilde{\phi}_c=
\tilde{\phi}_{\sigma\sigma}+\tilde{\phi}_{\sigma-\sigma} \ll 1$.
%\end{equation}
For such fillings, 
%\begin{equation}
$\tilde{\phi}_\gamma \gg 1$
%\end{equation}
so that
%\begin{equation}
$\tilde{\phi}_s \simeq 2\tilde{\phi}_\gamma$ or 
$\tilde{W}_s \simeq 2$. 
%\end{equation}

The Green function in the renormalized SSA is divided into coherent and
incoherent parts:
%
%\begin{equation}
$G_\sigma^{(0)} (i \varepsilon_{n}, {\bf k}) =
(1/\tilde{\phi}_\gamma)
g_\sigma^{(0)} (i \varepsilon_{n}, {\bf k})
\!+\! \mbox{(incoherent part)} $, 
%\end{equation}
%
with
\begin{equation}\label{EqGreenCoh}
g_\sigma^{(0)} (i \varepsilon_{n}, {\bf k}) =
\frac1{i \varepsilon_{n} +\mu^* 
- \xi ({\bf k}) + i \gamma \mbox{sign}(\varepsilon_{n})  } ,
\end{equation}
where $\mu^* =(\mu-\tilde{\Sigma}_0)/\tilde{\phi}_\gamma$ is an effective
chemical potential, 
$\xi ({\bf k}) =[E({\bf k})+\Delta\Sigma({\bf k})]/\tilde{\phi}_\gamma$
is the dispersion relation of quasiparticles in the renormalized SSA, and
$\mbox{sign}(\varepsilon_{n})=\varepsilon_{n}/|\varepsilon_{n}|$; the
incoherent part describes the lower and upper Hubbard bands. We introduce
a phenomenological life-time width $\gamma$, which is partly due to
disorder and partly due to many-body effects. Although $\gamma$ depends
on energies in general even if it is due to disorder, as is discussed in
Appendix~\ref{SecDisorder}, its energy dependence is
ignored.\cite{ComGamma} Effects of life-time widths or the coherence of
quasiparticles on the crossover between local-moment magnetism and
itinerant-electron magnetism can be, at least qualitatively, examined
even in this simplified scheme.

According to the Fermi-surface sum rule,\cite{Luttinger1,Luttinger2}
the number of electrons is given by that of quasiparticles;
the density or the number of electrons per site  for $T/T_K \rightarrow
+0$ and 
$\gamma/k_BT_K \rightarrow +0$ is given by
\begin{eqnarray}\label{EqFSumRule}
n &=& 
2\frac{k_BT}{N}\sum_{\varepsilon_n{\bf k}}
e^{i \varepsilon_{n}0^+} 
g_\sigma (i \varepsilon_{n}, {\bf k})
\nonumber \\ &=& 
2 \!\! \int \!\! d\varepsilon \rho_{\gamma\rightarrow 0}(\varepsilon) 
f_\gamma(\varepsilon \!-\! \mu^*) 
\nonumber \\ &=&
2 \!\!\int \! d\varepsilon \rho_{\gamma}(\varepsilon) 
f_{\gamma=0}(\varepsilon \!-\! \mu^*) ,
\end{eqnarray}
with
\begin{equation}
\rho_\gamma (\varepsilon) = \frac1{\pi N} \sum_{\bf k}
\frac{\gamma}
{\left[\varepsilon - \xi({\bf k})\right]^2 +\gamma^2} , \qquad 
\end{equation}
\begin{equation}\label{EqfGamma}
f_\gamma(\varepsilon)  =
\frac1{2} +\frac1{\pi} \mbox{Im}  \left[
\psi\left( \frac1{2} 
+ \frac{\gamma - i\varepsilon}{2\pi k_B T} \right)
\right] ,
\end{equation}
with $\psi (z)$ the di-gamma function. Note that
%\begin{equation}
$f_{\gamma=0}(\varepsilon) =1/\left[e^{\varepsilon/k_BT}+1\right] $.
%\end{equation}
We assume Eq.~(\ref{EqFSumRule}) even for nonzero 
$T$ and $\gamma$. The parameter $\tilde{\Sigma}_0$
or $\mu^*$ can be determined from  Eq.~(\ref{EqFSumRule}) as a function
of $n$.

\subsection{Intersite exchange interaction}

Denote susceptibilities of the Anderson and the $t$-$J$ models, which
do not include the factor $\frac1{4}g^2\mu_B^2$ with $g$ being the $g$
factor and $\mu_B$ the Bohr magneton, by 
$\tilde{\chi}_s(i\omega_l)$ and $\chi_s(i\omega_l, {\bf q})$,
respectively. In Kondo lattices, local spin fluctuations at different
sites interact with each other by an exchange interaction.  Following
this physical picture, we define an exchange interaction
$I_s(i\omega_l, {\bf q})$ by 
\begin{equation}\label{EqKondoSus}
\chi_s(i\omega_l, {\bf q}) =
\frac{\tilde{\chi}_s(i\omega_l)}
{1 - \mbox{$\frac{1}{4}$}I_s(i\omega_l, {\bf q})
\tilde{\chi}_s(i\omega_l)} . 
\end{equation}
Following the previous paper,\cite{OhkawaPseudogap} we obtain 
\begin{equation}\label{EqIs}
I_s (i\omega_l, {\bf q}) = J({\bf q}) + 
2 U_\infty^2 \Delta\pi_s(i\omega_l, {\bf q}) ,
\end{equation}
with $\Delta\pi_s(i\omega_l, {\bf q})$ the multi-site part of the
irreducible polarization function in spin channels; 
$2 U_\infty^2 \Delta\pi_s(i\omega_l, {\bf q})$ is examined
in Secs.~\ref{SecExc} and \ref{SecMode}.

When the Ward relation \cite{ward} is made use of,
the irreducible single-site three-point vertex function in spin channels, 
$\tilde{\lambda}_s(i\varepsilon_n,i\varepsilon_n \!+\! i\omega_l;i\omega_l)$,
is given by
\begin{equation}\label{EqThreeL}
 U_\infty 
\tilde{\lambda}_s(i\varepsilon_n,i\varepsilon_n+i\omega_l;i\omega_l)
= 2\tilde{\phi}_s / \tilde{\chi}_s(i\omega_l)  ,
\end{equation}
for $|\varepsilon_n| \rightarrow +0$ and 
$|\omega_l| \rightarrow +0$.
 We approximately use Eq.~(\ref{EqThreeL}) for 
$|\varepsilon_n| \alt 2 k_BT_K$ and
$|\omega_l| \alt 2 k_BT_K$, with $T_K$ the Kondo temperature defined by
\begin{equation}\label{EqDefTK} 
k_BT_K = \left[1/\tilde{\chi}_s(0)
\right]_{T\rightarrow 0}. 
\end{equation} 
An exchange interaction mediated by spin fluctuations is calculated in
such a way that
%%%%%%%%%%%%%%%%%%%%%%%%%%%%%%%%%%%%%%%%%%%%%%%%%%%%%%%%%%%%%%%%%%%
\begin{equation}\label{SpinFluctI}
\frac{1}{4} \!\left[
2\tilde{\phi}_s / \tilde{\chi}_s(i\omega_l)\right]^2 
 F_s(i\omega_l,{\bf q}) = 
\tilde{\phi}_s^2 \frac{1}{4} I_s^* (i\omega_l, {\bf q}) ,
\end{equation}
%%%%%%%%%%%%%%%%%%%%%%%%%%%%%%%%%%%%%%%%%%%%%%%%%%%%%%%%%%%%%%%%%%%
with 
\begin{eqnarray}\label{EqF}
 F_s(i\omega_l,{\bf q}) &=& 
\chi_s(i\omega_l,{\bf q}) - \tilde{\chi}_s(i\omega_l) ,
\\ 
\label{EqIs*2}
\frac{1}{4} I_s^*(i\omega_l, {\bf q}) &=&
\frac{ \frac{1}{4}
I_s (i\omega_l, {\bf q}) }
{1 - \frac{1}{4}I_s(i\omega_l, {\bf q})
\tilde{\chi}_s(i\omega_l) } .
\end{eqnarray}
The single-site term is subtracted in $ F_s(i\omega_l,{\bf q})$ because it
is considered in SSA. 

Because of these equations, we call
$I_s(i\omega_l, {\bf q})$ a  {\em bare} exchange interaction,
$I_s^*(i\omega_l, {\bf q})$ an enhanced one, and $\tilde{\phi}_s$ an
effective three-point vertex function in spin channels. Because the
spin space is isotropic,  the interaction in the transversal channels
is also given by these equations. Intersite effects can be
perturbatively considered in terms of $I_s(i\omega_l, {\bf q})$,
$I_s^*(i\omega_l, {\bf q})$ or $F_s(i\omega_l,{\bf q})$ depending on
each situation.\cite{CommentSC}

%%%%%%%%%%%%%%%%%%%%%%%%%%%%%%%%%%%%%%%%%%%%%%%%%%%%%
%%%%%%%%%%%%%%%%%%%%%%%%%%%%%%%%%%%%%%%%%%%%%%%%%%%%% 
\begin{figure*}
\centerline{\hspace*{0.8cm}
\includegraphics[width=4.8cm]{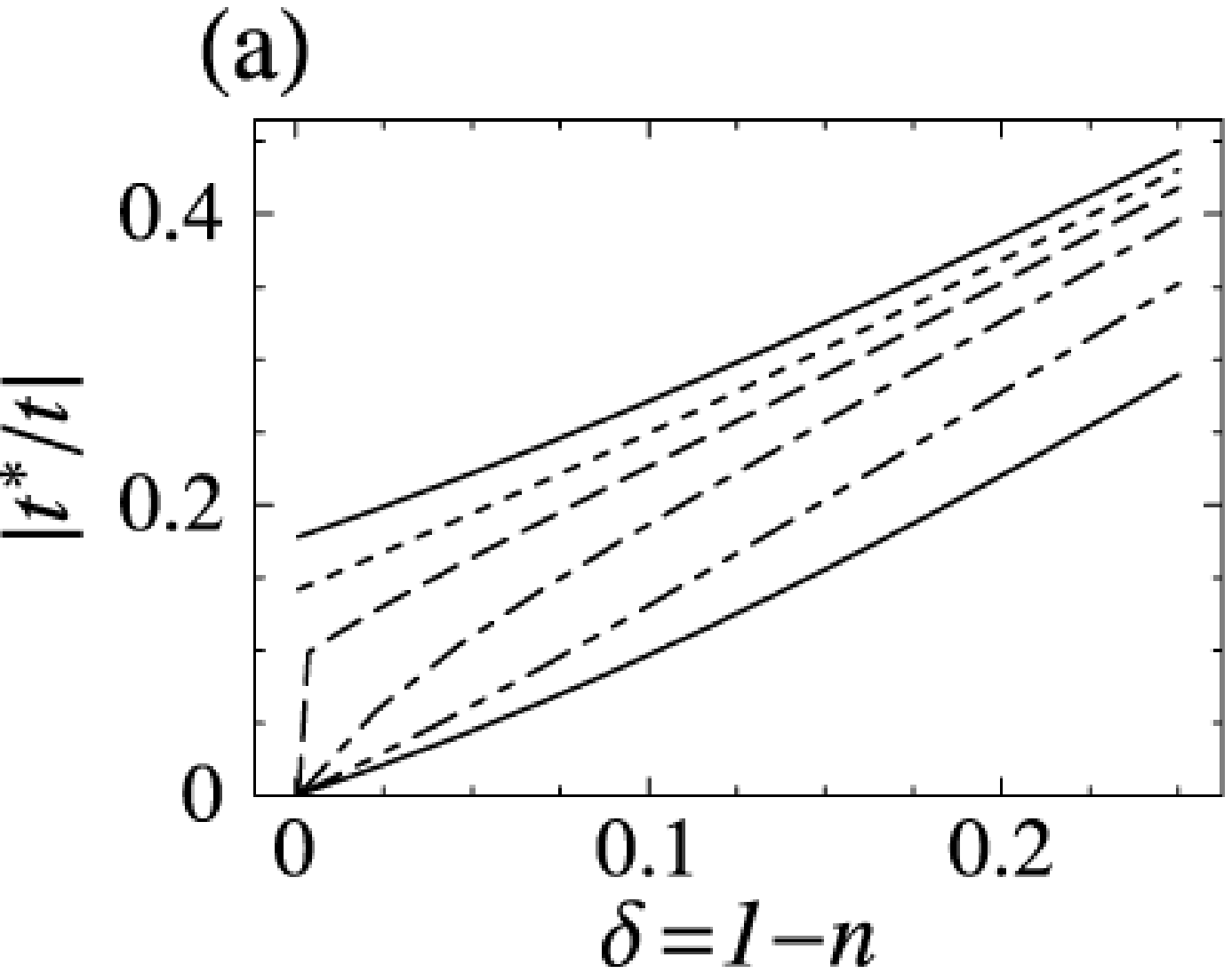}\hspace{-0.3cm}%
\includegraphics[width=4.8cm]{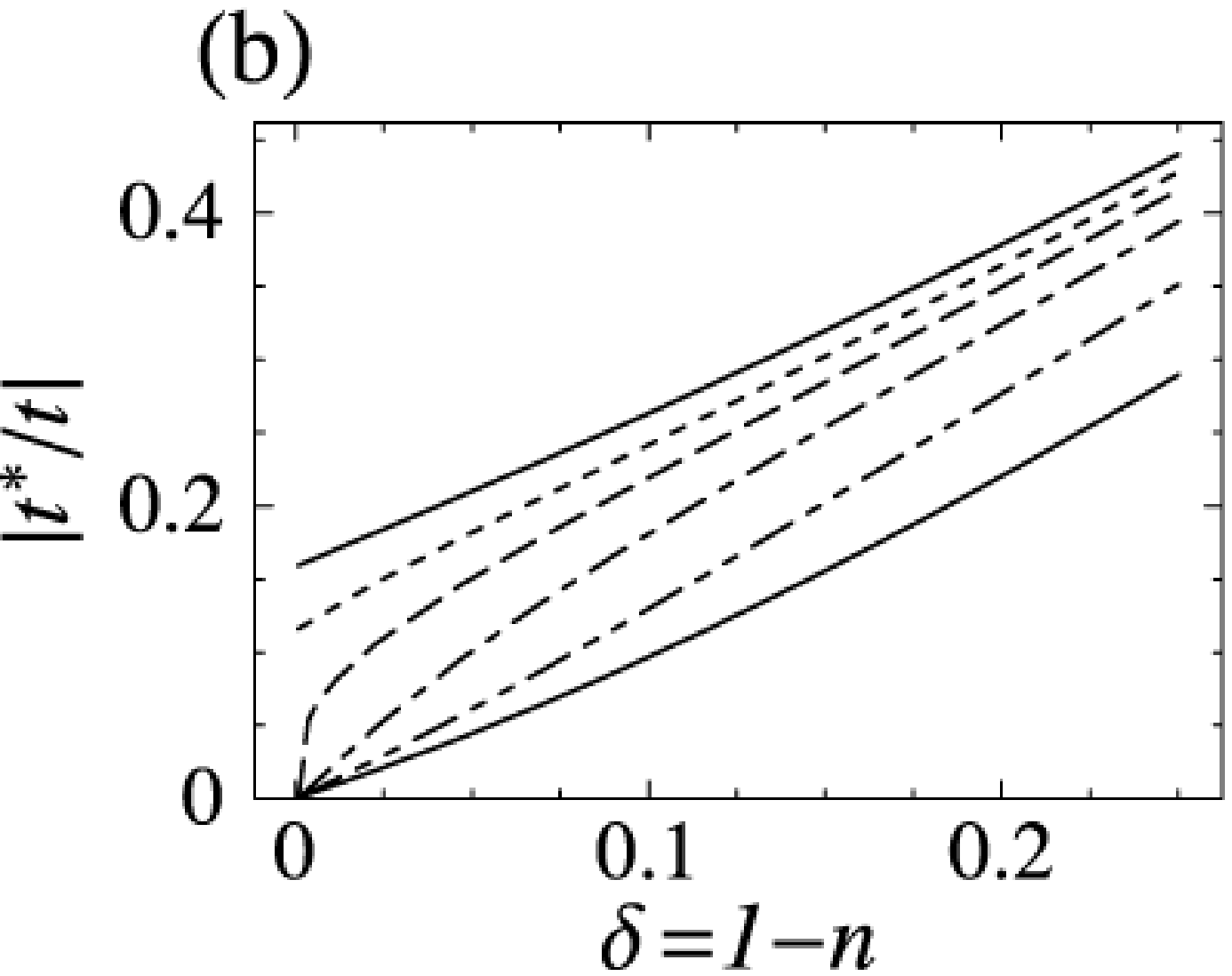}\hspace{-0.3cm}%
\includegraphics[width=4.8cm]{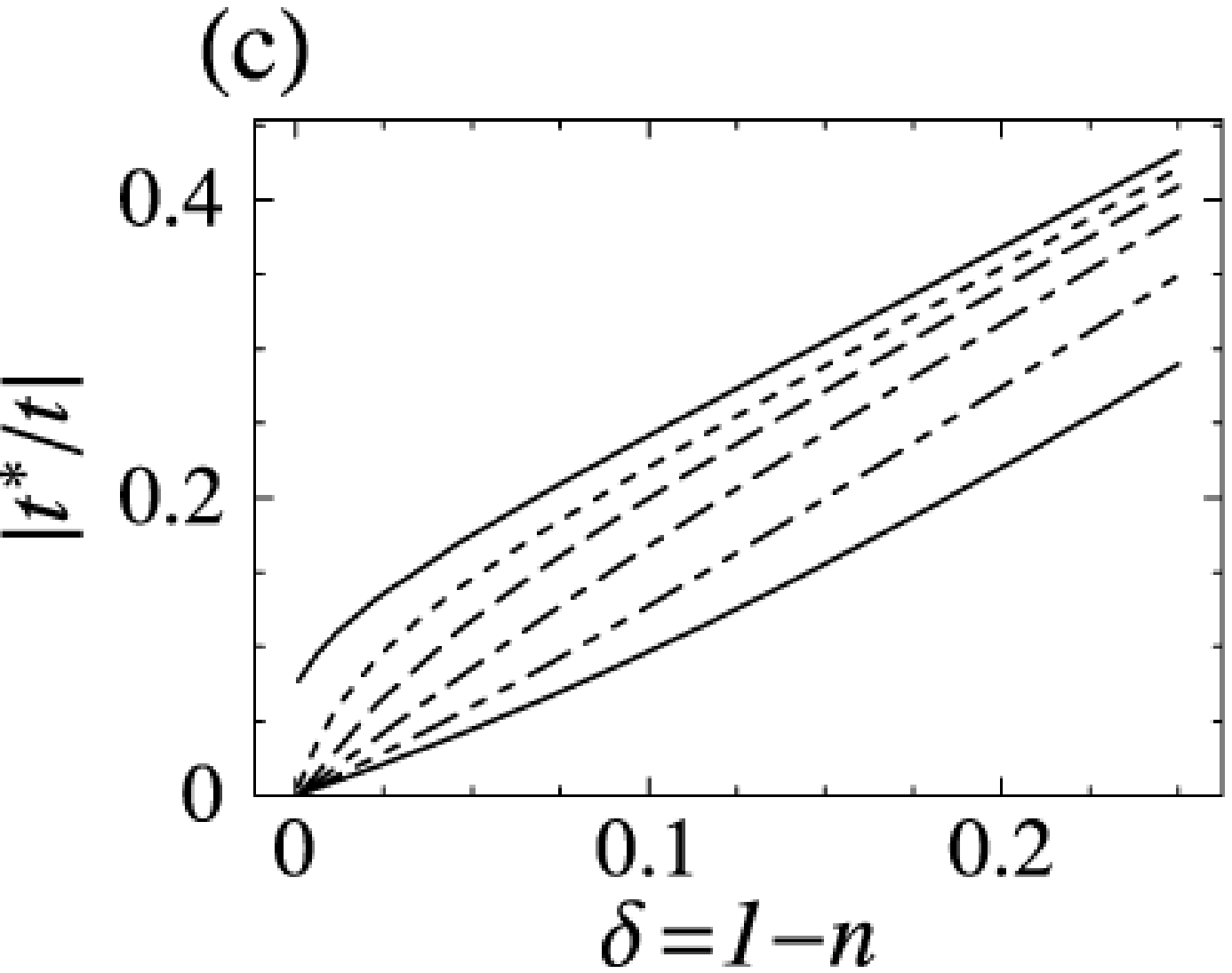}\hspace{-0.3cm}%
\includegraphics[width=4.8cm]{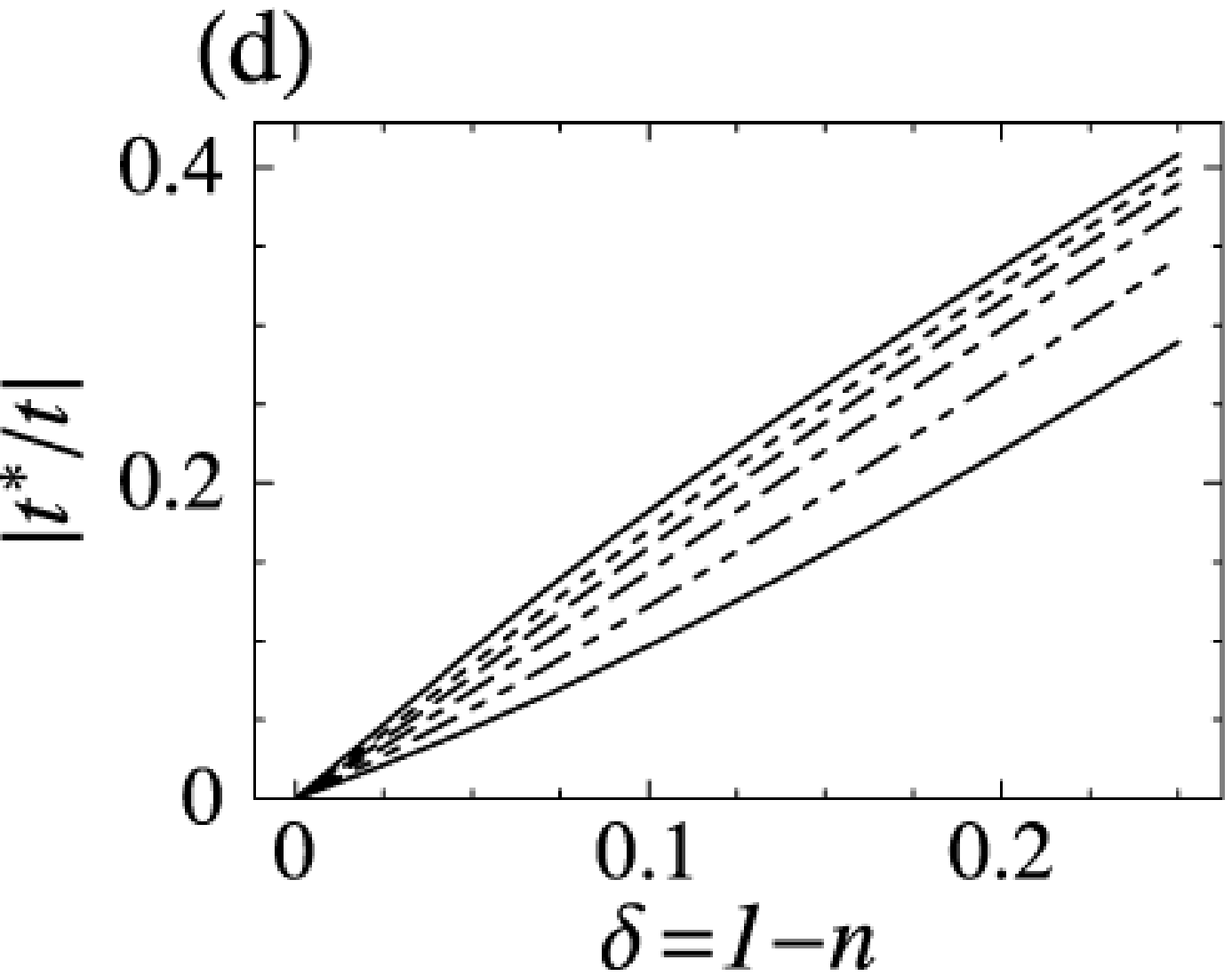}
}
\caption[1]{
 Renormalized transfer integrals $t^*$ of quasiparticles
in the unperturbed state of the symmetric model:    
(a) $k_BT/|t|=0.02$, (b) $k_BT/|t|=0.1$,
(c) $k_BT/|t|=0.2$, and
(d) $k_BT/|t|=0.4$.
In each figure, topmost solid, dotted, broken,
dot-broken, and two-dot-broken lines show results for
$\gamma/|t|=0.01$,  0.1, 0.2, and 0.4, respectively.  
For comparison,
$1/\tilde{\phi}_\gamma$ is also shown by a bottom solid line.
}
\label{t^star}
\end{figure*}
%%%%%%%%%%%%%%%%%%%%%%%%%%%%%%%%%%%%%%%%%%%%%%%%%%%%%
%%%%%%%%%%%%%%%%%%%%%%%%%%%%%%%%%%%%%%%%%%%%%%%%%%%%%
\begin{figure}
\includegraphics[width=8cm]{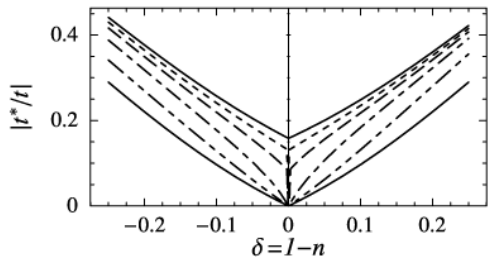}
\caption[2]{
$t^*$ of the asymmetric model for $k_BT/|t|=0.02$.  See also the
caption of Fig.~\ref{t^star}; this figure corresponds to
Fig.~\ref{t^star}(a) for the symmetric model.  
 % More than half filling case or electron doping cases $(n > 1$ or
 % $\delta<0)$ can be treated when we take the hole picture. 
 % For small $\gamma$, $|t^*|$ is relatively a little larger in
 % electron doping cases than it is in hole doping cases.
 }
\label{t^star-asym}
\end{figure}
%%%%%%%%%%%%%%%%%%%%%%%%%%%%%%%%%%%%%%%%%%%%%%%%%%%%%
%%%%%%%%%%%%%%%%%%%%%%%%%%%%%%%%%%%%%%%%%%%%%%%%%%%%%

\subsection{Fock term of the superexchange interaction}
\label{SecFock}

Note that $ \lim_{\omega_l \rightarrow +\infty}
 I_s^*(i\omega_l, {\bf q})  = J({\bf q}) $.
We consider the multisite selfenergy correction due
to high-energy spin excitations or due to the superexchange
interaction $J({\bf q})$. The Fock term of $J({\bf q})$  gives a
selfenergy correction independent of energies:\cite{exchByOhkawa}
\begin{equation}\label{EqSelf-1}
\Delta \Sigma_\sigma ({\bf k}) \!=\!
\frac{3}{4} \tilde{\phi}_s^2
\frac{k_B T}{N} \!\!
\sum_{\varepsilon_{n}{\bf p}}
\! J({\bf k}\!-\!{\bf p}) 
e^{i \varepsilon_{m}0^+}\!
G_\sigma (i \varepsilon_{n}, {\bf p}) .
\end{equation}
The factor 3 appears because of three
spin channels.  When
only the coherent part is considered,  
\begin{equation}
\frac1{\tilde{\phi}_\gamma}\Delta \Sigma_\sigma ({\bf k}) =
\frac{3}{4} \tilde{W}_s^2
J \Xi \eta_{1s}({\bf k})  , 
\end{equation}
with
\begin{equation}\label{EqXi}
\Xi =
\frac1{ N}\sum_{\bf k} \eta_{s} ({\bf k}) 
f_\gamma\left[\xi({\bf k})-\mu^* \right] .
\end{equation}
%%%%%%%%%%%%%%%%%%%%%%%%%%%%%%%%%%%%%%%%%%%%%%%%%%%%%%%%%%%%%%%%%%%%%%%%%%
The dispersion relation of quasiparticles is given
by   
\begin{equation}
\xi({\bf k}) = 
- 2t^* \eta_{1s}({\bf k}) 
- 2t_2^* \eta_{2s}({\bf k})
\end{equation} 
in the renormalized SSA.
The effective transfer integral $t^*$  should be
selfconsistently determined to satisfy
\begin{equation}\label{Eqt*}
2 t^* = \frac{2t}{\tilde{\phi}_\gamma} 
- \frac{3}{4}\tilde{W}_s^2 J \Xi, 
\end{equation}
while $t_2^*$ is simply given by
$t_2^* = t^\prime/\tilde{\phi}_\gamma $.

For the symmetric model, $t^\prime=0$ so that $t_2^*=0$. 
In order to examine how crucial role the shape of the Fermi
surface plays in the asymmetry, we consider a phenomenological
asymmetric model with
\begin{equation}
t_2^*/t^* = - 0.3 . 
\end{equation}

Expansion parameters $\tilde{\phi}_\gamma$ and $\tilde{\phi}_{s}$ 
are given by those of  the mapped Anderson model, which should be
selfconsistently determined to satisfy Eqs.~(\ref{EqMap}) and
(\ref{Eqt*}).  However, we approximately use those for the Anderson model
with a constant hybridization energy. According to Appendix of the
previous paper,\cite{ferromagnetism}
\begin{equation}\label{EqPhiG}
\tilde{\phi}_\gamma= \frac1{2}
\left(\frac{1}{|\delta|} + |\delta|\right)
\frac{\displaystyle \left(\pi /2 \right)^2(1-|\delta|)^2}
{\displaystyle \cos^2 \!\left(\pi \delta/2 \right)} ,
\end{equation}
%and 
\begin{equation}\label{EqPhiS}
\tilde{\phi}_{s} = \frac{1}{|\delta|} 
\frac{\displaystyle \left(\pi /2 \right)^2(1-|\delta|)^2}
{\displaystyle \cos^2 \!\left(\pi \delta/2 \right)} ,
\end{equation}
where 
\begin{equation}
\delta = 1 - n
\end{equation}
is the concentration of dopants, {\it holes}  ($\delta>0$) or
electrons ($\delta<0$). These are consistent with 
Gutzwiller's theory.\cite{Gutzwiller}
%%%%%%%%%%%%%%%%%%%%%%%%%%%%%%%%%%%%%%%%%%%%%%%%%%%%%%%%%%%%%%%%%%

Figures~\ref{t^star} and \ref{t^star-asym} show $t^*$ of the symmetric and
asymmetric models, respectively,  as a function of  $\delta$.  
 It is interesting that $t^*$
is nonzero even for $\delta\rightarrow 0$ if life-time widths $\gamma$
are small enough and temperatures $T$ are low enough. 
For the symmetric model ($t_2^* = 0$), 
 Eq.~(\ref{EqXi}) can be analytically calculated for 
$T =0$~K, $\gamma =0$ and  $\delta =0$ $(\mu^* =0)$ so that
$\Xi=4/\pi^2 $. Then, $\left[t^*/t\right]_{\delta \rightarrow 0}
\rightarrow -3\tilde{W}_s^2 (J/t)/2\pi^2 = 0.18$ for
the symmetric model. If
$\gamma$  are large enough or $T$ are high enough, on the other hand, 
$\Xi$ and $t^*$ vanish for $\delta\rightarrow 0$.

\begin{figure*}
\centerline{\hspace*{0.5cm}
\includegraphics[width=5.6cm]{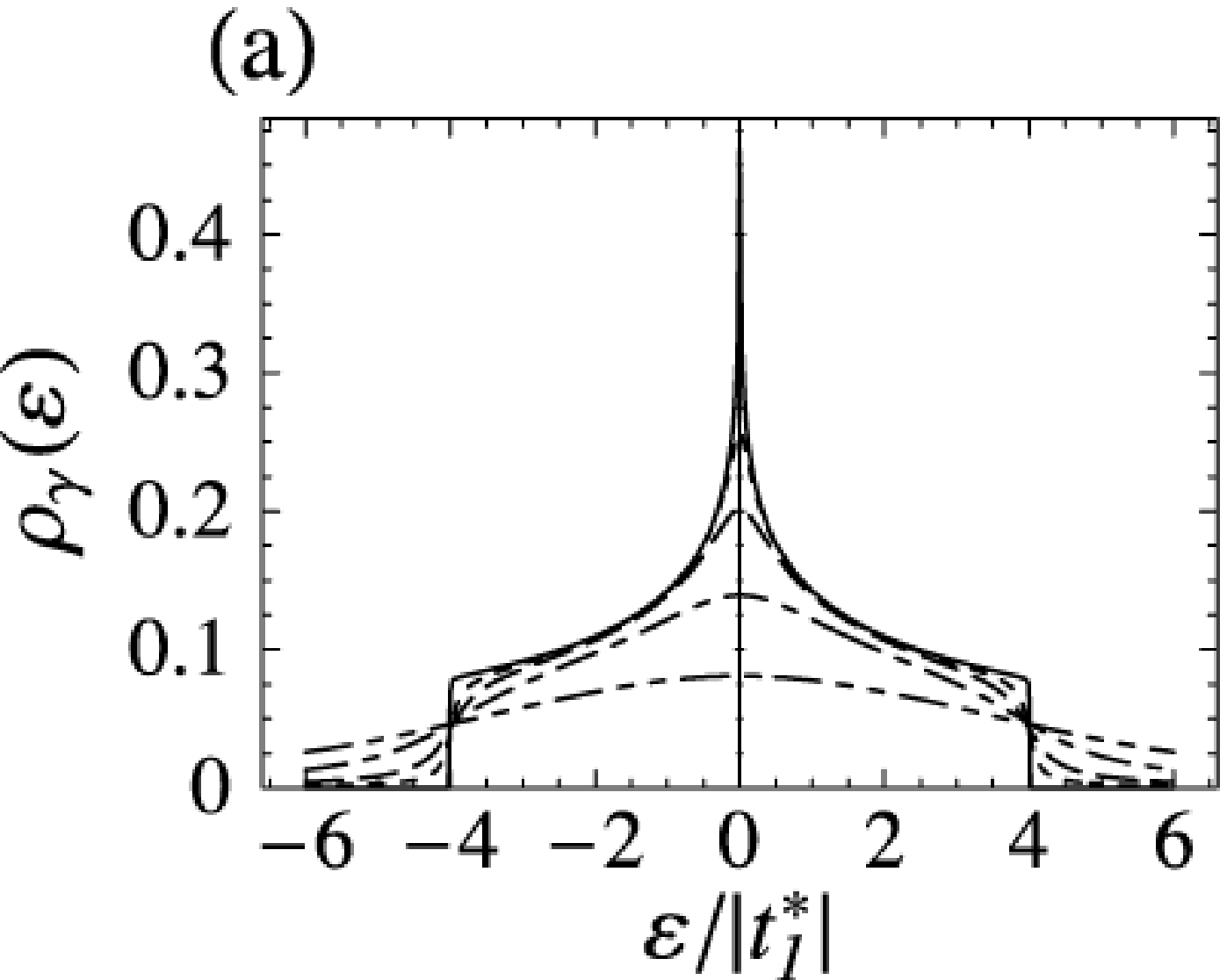}\hspace{-0.9cm}%
\includegraphics[width=5.0cm]{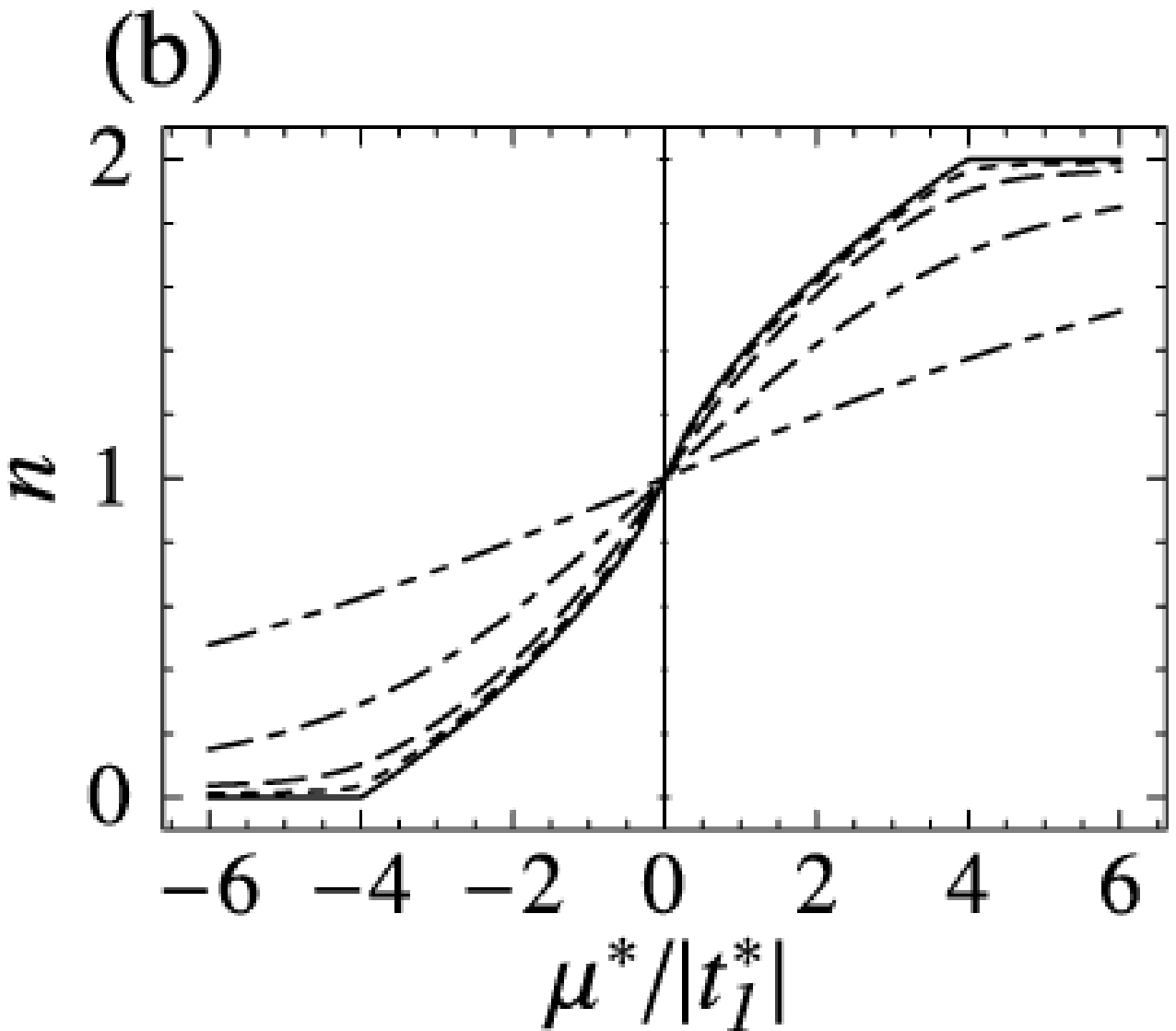} \hspace{-0.6cm}%
\includegraphics[width=5.1cm]{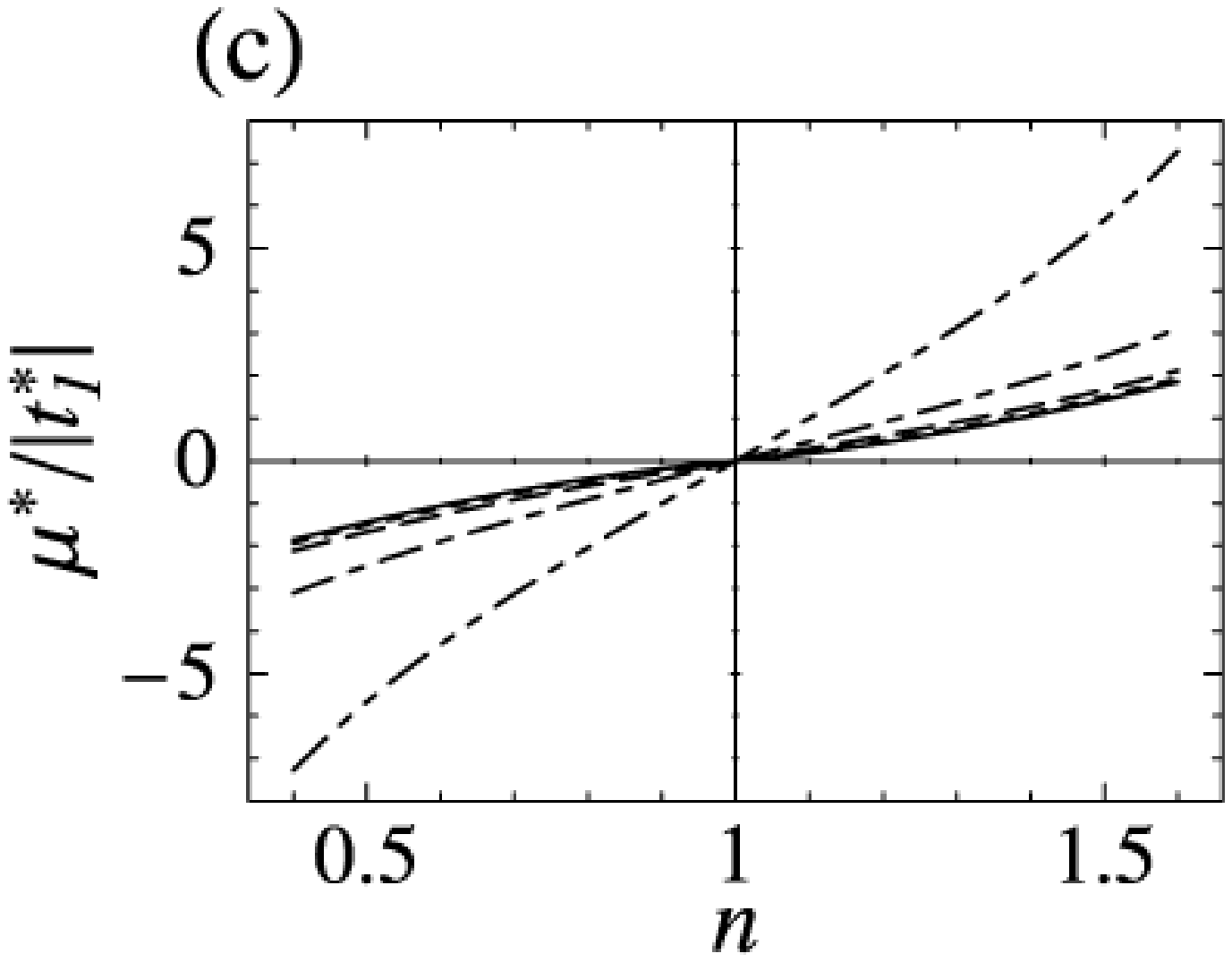} \hspace{-0.5cm}%
\includegraphics[width=4.3cm]{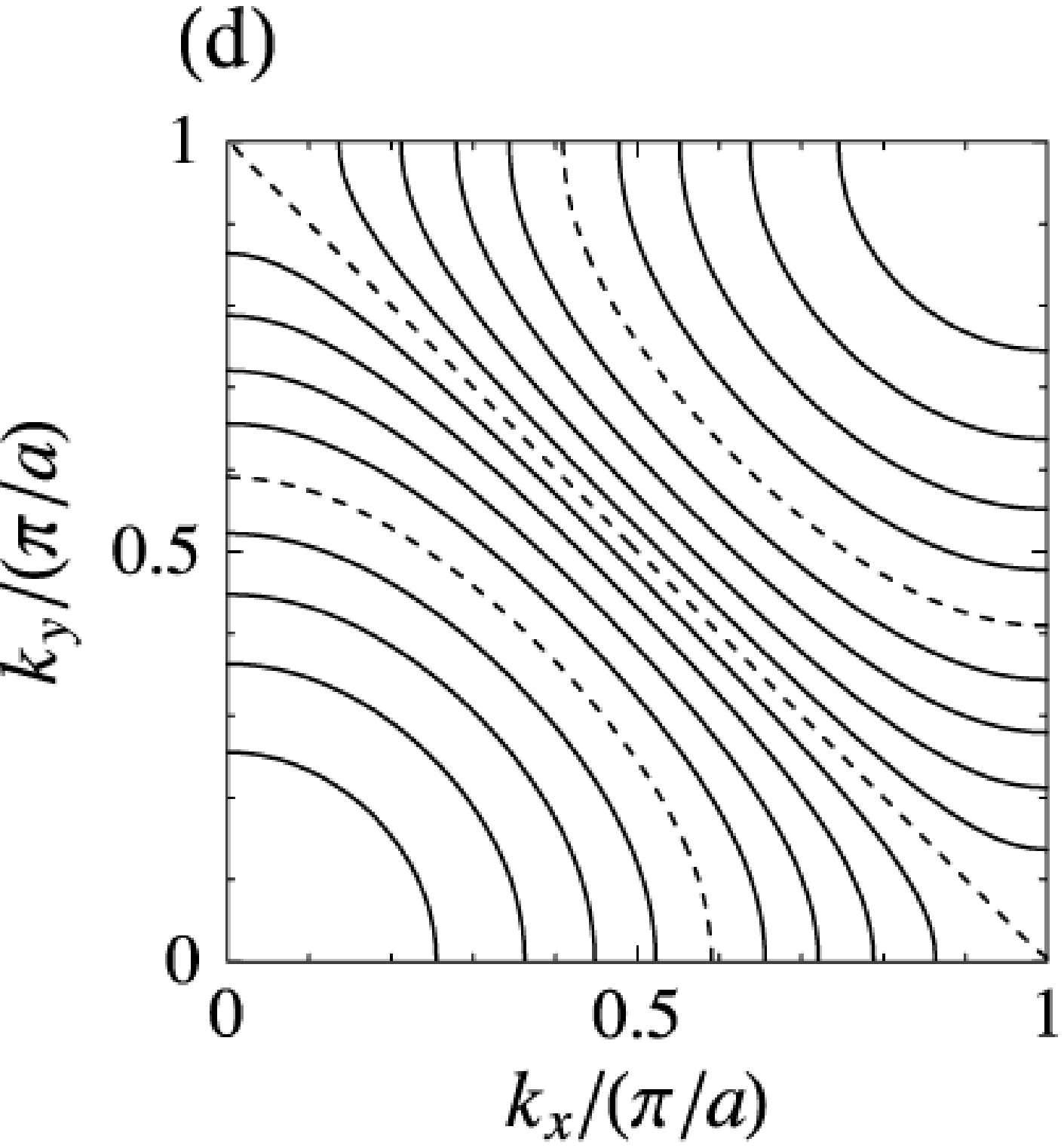}
}
\caption[3]{
Single-particle properties of the unperturbed state of the
symmetric model.   (a)
Density of states for quasiparticles
$\rho_\gamma(\varepsilon)$,  
(b) effective chemical potentials, $\mu^*$, as functions of
carrier concentrations $n$, and
(c) $n$ as functions of $\mu^*$. 
In these three figures, solid, dotted, broken, dot-broken, and
two-dot-broken lines show results for
$\gamma/|t^*|=10^{-3}$,  0.1, 0.3, 1, and 3, respectively.
 Note that the solid and dotted lines are the
almost same as each other. (d) Fermi surfaces for $k_BT=0$,
$\gamma=0$, and 19 electron concentrations such as  $n=0.1 \times
i$, with $1\le i \le 19$ being an integer. Dotted lines show
Fermi surfaces for  $n=$0.5, 1.0, and 1.5. Single-particle
properties of the asymmetric model can be found in Fig.~2 of
Ref.~\onlinecite{OhkawaPseudogap}. }
\label{rho_n_mu_FS}
\end{figure*}

Figure~\ref{rho_n_mu_FS} shows physical properties of the  unperturbed
state of the symmetric model:  $\rho_\gamma(\varepsilon)$, 
$\mu^*$ as a function of $n$, $n$ as a function of $\mu^*$, and Fermi
surfaces for various $n$. Physical properties of the unperturbed state of
the asymmetric model can be found in Fig.~2 of
Ref.~\onlinecite{OhkawaPseudogap}.

%%%%%%%%%%%%%%%%%%%%%%%%%%%%%%%%%%%%%%%%%%%%%%%%%%%%%
%%%%%%%%%%%%%%%%%%%%%%%%%%%%%%%%%%%%%%%%%%%%%%%%%%%%%
%%%%%%%%%%%%%%%%%%%%%%%%%%%%%%%%%%%%%%%%%%%%%%%%%%%%%
%%%%%%%%%%%%%%%%%%%%%%%%%%%%%%%%%%%%%%%%%%%%%%%%%%%%%

\section{Suppression of the N\'{e}el temperature by spin fluctuations} 
\label{SecTN}

\subsection{Renormalization of the Kondo temperature}

% We consider an instability condition of the unperturbed
% state constructed in the renormalized SSA against antiferromagnetism.
 According to Eq.~(\ref{EqKondoSus}), 
the N\'{e}el temperature $T_N$ is determined by 
\begin{equation}\label{EqAFCondition}
1 - \frac1{4} I_s(0, {\bf Q})\tilde{\chi}_s(0) = 0.
\end{equation}
Here, {\bf Q}  is an
ordering wave number to be determined. 

The local susceptibility $\tilde{\chi}_s(0)$ is almost constant at 
$T \ll T_K$, while it obeys the Curie law at high
temperatures such as 
$\tilde{\chi}_s(0)= n/k_BT$ at $T\gg T_K$. 
In this paper,
we use an interpolation between the two limits:
\begin{equation}\label{EqCW-local}
\tilde{\chi}_s(0) = \frac{n}
{\displaystyle k_B \sqrt{n^2T_K^2+ T^2}} ,
\end{equation}
with $T_K$ defined by Eq.~(\ref{EqDefTK}).

The renormalization of $t^*$ by the Fock term is nothing but the
renormalization of local quantum spin fluctuations or their energy
scale $k_BT_K$  by the superexchange interaction. According to
Eq.~(\ref{EqDefTK}) together with the Fermi-liquid 
relation\cite{Yamada,Yosida} and the mapping condition (\ref{EqMap}), the
static susceptibility or $k_BT_K$ is given by
\begin{eqnarray}\label{EqTK2}
\left[\tilde{\chi}_s(0)\right]_{T=0\mbox{~\scriptsize K}}   &=&
\frac1{k_BT_K}
 % 2 \tilde{\phi}_s \frac1{N}\sum_{\bf k}
 %  \left( - \frac1{\pi} \right) \mbox{Im}  
 %  \left[ G_\sigma^{(0)} (+i0,{\bf k}) \right]_{T\rightarrow+0}  
 %
 % \nonumber \\    &=&
= 2 \tilde{W}_s \left[\rho_{\gamma} (\mu^*)
\right]_{\gamma \rightarrow 0}, 
\end{eqnarray}
in the absence of disorder. In disordered systems, the
mapping conditions are different from site to site so that $T_K$ are also
different from site to site. Such disorder in $T_K$ causes 
energy-dependent life-time width, as is studied in
Appendix~\ref{SecDisorder}. However, life-time widths due to the
disorder in $T_K$ are small on the chemical potential in case of
non-magnetic impurities. Then, a mean value of $T_K$ in disordered systems
is approximately given by Eq.~(\ref{EqTK2}) with nonzero but
small $\gamma$.  It follows from Eq.~(\ref{EqTK2}) that
\begin{equation}\label{EqTKdef}
k_BT_K = \frac{|t^*|}{c_{T_K}}\frac{ 1}{2\tilde{W}_s} ,
\end{equation}
with $c_{T_K}$  a numerical constant depending on $n$. 
As is shown in Fig.~\ref{rho_n_mu_FS}(a), 
$\rho_\gamma(\mu^*) \simeq 0.15$ for 
$0.1 \alt \gamma/|t^*| \alt 1$ and
$0.05 \alt |\delta| \alt 0.25$.
We assume that  $c_{T_K}$ is independent of $n$ for the sake of
simplicity:  %%%%%%%%%%%%%%%%%%%%%%%%%%%%%%%%%%%%%%%%%%%\cite{ComCTK} 
\begin{equation}\label{EqCTK}
c_{T_K}=0.15 .
\end{equation}
We are only interested in physical properties that never drastically
change when  $c_{T_K}$ slightly changes.

\subsection{Exchange interaction arising from the virtual exchange of
pair excitations of quasiparticles}
\label{SecExc}

\begin{figure*}
\centerline{\hspace*{0.7cm}
\includegraphics[width=5.8cm]{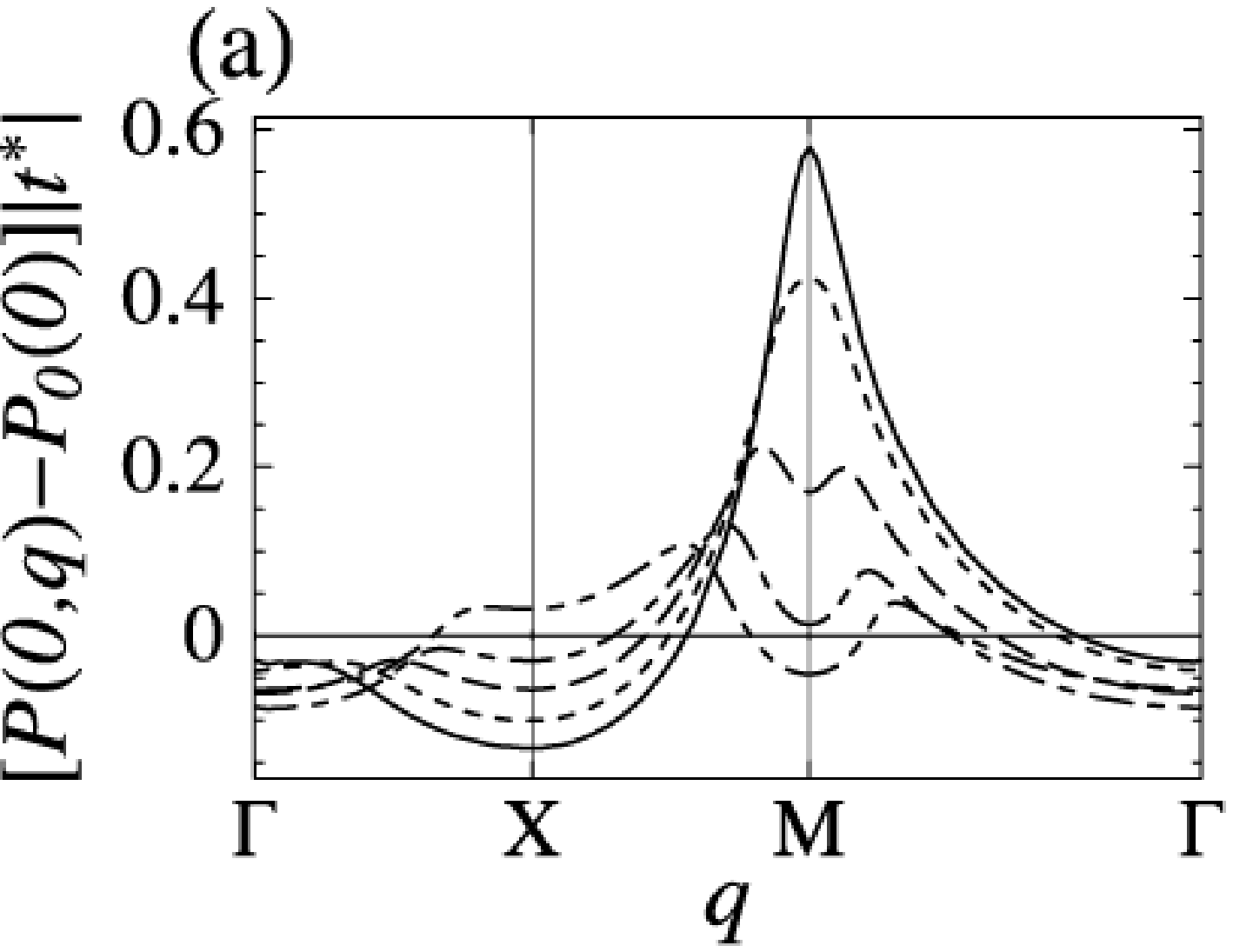}
\includegraphics[width=5.8cm]{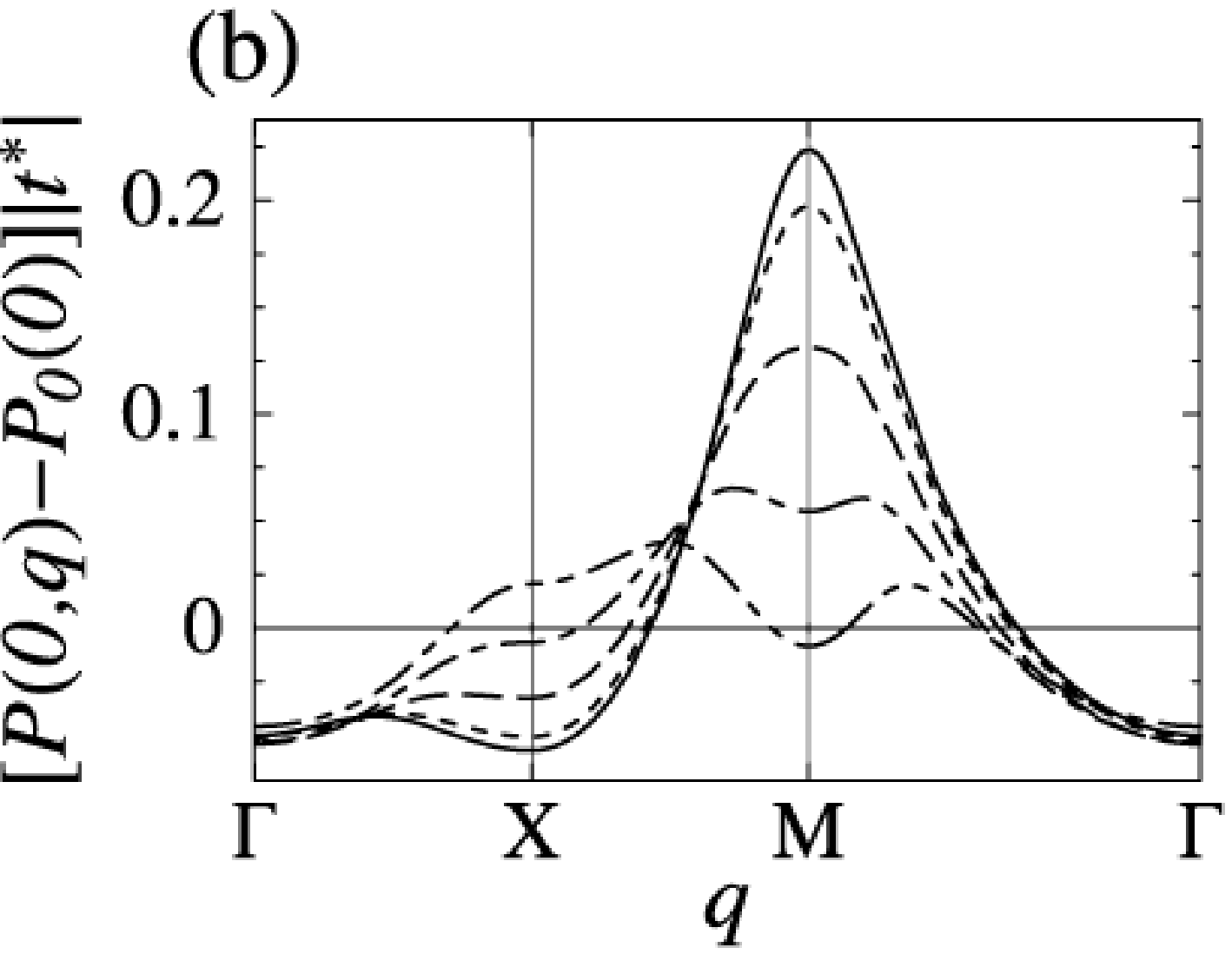}
\includegraphics[width=6.2cm]{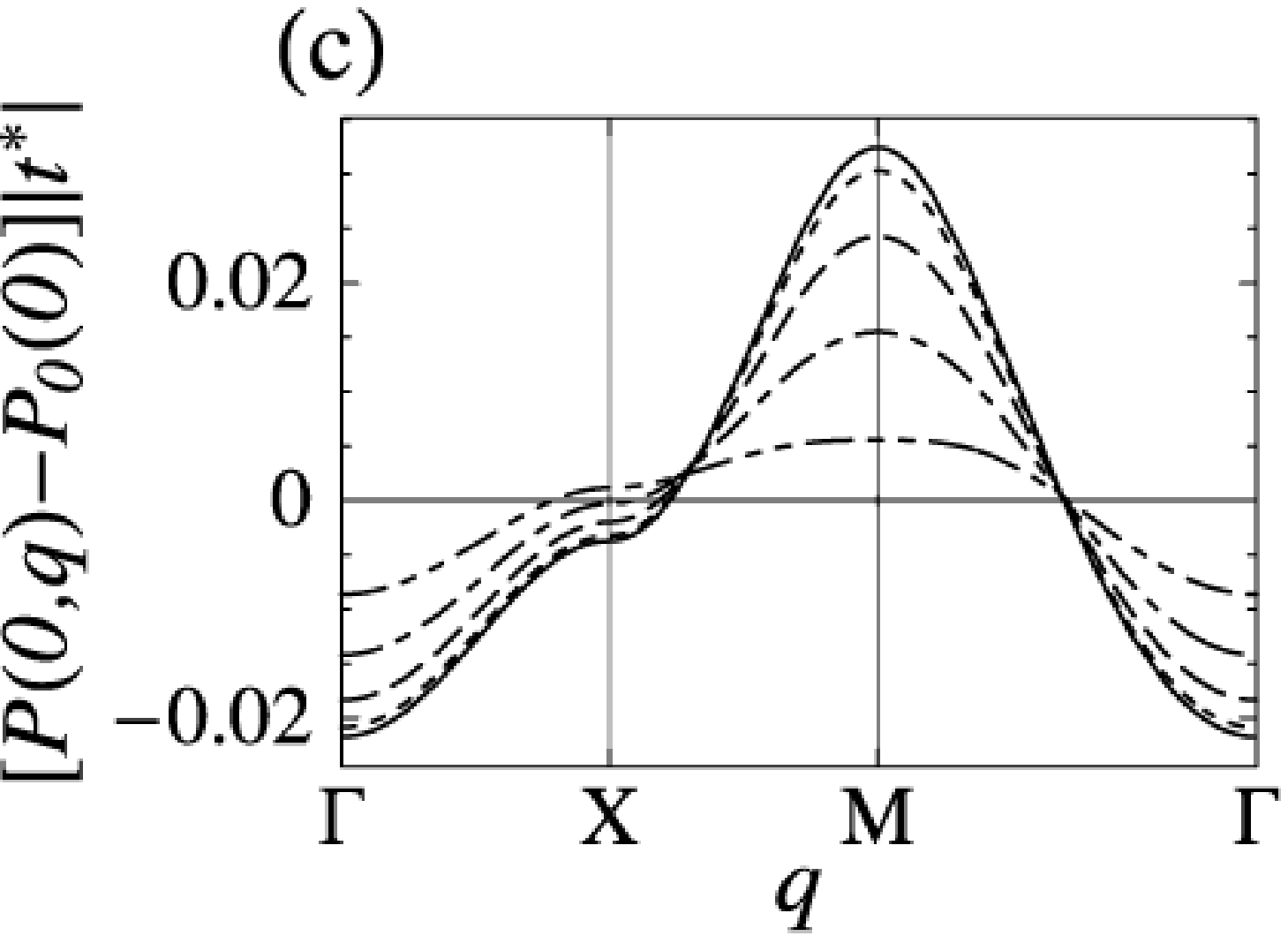}
}
\caption[4]{ 
Static polarization function
$[P(0,{\bf q})-P_0(0)]|t^*|$ of the symmetric model:
(a) $k_BT/|t^*|=\gamma/|t^*|=0.1$,
(b) $k_BT/|t^*|=\gamma/|t^*|=0.3$, and
(c) $k_BT/|t^*|=\gamma/|t^*|=1$.
Solid, dotted, broken, dot-broken, and double-dot-broken lines show results
for
$n \rightarrow 1$, $n=0.9$, 0.8, 0.7, and 0.6, respectively.
Here, $\Gamma$, $X$ and $M$ stand for $(0,0)$, $(\pi/a,0)$ and
$(\pi/a,\pi/a)$, respectively.
}
\label{pol-sym}
\end{figure*}

\begin{figure} 
\centerline{\hspace*{0.8cm}
\includegraphics[width=6.5cm]{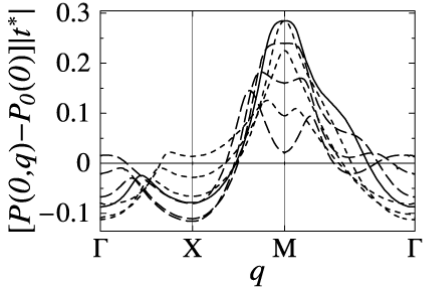}
}
\caption[5]{ 
$[P(0,{\bf q})-P_0(0)]|t^*|$ of the asymmetric model:
$k_BT/|t^*|=\gamma/|t^*|=0.1$.
A solid line shows a result for $n \rightarrow 1$, dotted lines results
of electron doping cases such as $n=1.1$, 1.2 and 1.3, and  broken
lines  results for hole doping cases such as $n=0.9$, 0.8,  and
0.7. In either doping case, the polarization function at $M$ point
decreases with increasing concentrations of dopants $|\delta|=|1-n|$.  
 % Note that the polarization function at M point is relatively larger in
 % electron doping cases than it is in hole doping cases. 
 }
\label{pol-asy}
\end{figure}

The first term of Eq.~(\ref{EqIs}) is the superexchange
interaction.\cite{exchByOhkawa} The second term is the sum of  an
exchange interaction arising from that of pair excitations of
quasiparticles, 
$J_Q(i\omega_l, {\bf q})$, and  the mode-mode coupling term,
$-4\Lambda (i\omega_l, {\bf q})$:
\begin{equation}
2 U_\infty^2 \Delta\pi_s(i\omega_l, {\bf q}) =
J_Q(i \omega_l, {\bf q}) - 4 \Lambda (i \omega_l, {\bf q}).
\end{equation}
 When higher-order terms in intersite effects are ignored,  
\begin{equation}\label{EqJQ1}
J_Q(i\omega_l, {\bf q}) =
4\left[\frac{\tilde{W}_s}{\tilde{\chi}_s(0) }\right]^2 
\left[ P(i\omega_l, {\bf q}) - P_0(i\omega_l)
\right], 
\end{equation}
with
\begin{equation}\label{EqP}
P(i \omega_l, {\bf q}) =
\frac{k_BT}{N} \!\!\sum_{\varepsilon_n{\bf k}\sigma} \!
g_\sigma^{(0)}(i\varepsilon_n \!+\! i\omega_l, {\bf k}
 \!+\! {\bf q})  g_\sigma^{(0)}(i\varepsilon_n, {\bf k}) .
\end{equation}
The local contribution 
$P_0(i\omega_l) =(1/N)\sum_{\bf q} P(i \omega_l, {\bf q})$ is
subtracted because it is considered in SSA. The static component is
simply given by
\begin{equation}
P(0,{\bf q}) =
\frac{2}{N} \!\sum_{\bf k}\! \frac{f_\gamma 
\bigl[\xi({\bf k} \!+\! {\bf q}) \!-\! \mu^*\bigr]
\!-\! f_\gamma \bigl[\xi({\bf k}) \!-\! \mu^*\bigr]}
{ \xi({\bf k}) -\xi({\bf k} +{\bf q}) -i0} .
\end{equation}
\normalsize 
Figures~\ref{pol-sym} and \ref{pol-asy} show $[P(0,{\bf q})-P_0(0)]$ of
the symmetric and asymmetric models. The polarization function is
relatively larger in electron-doping cases than it is in hole-doping
cases.

The magnitude of $J_Q(i\omega_l, {\bf q})$ is proportional to $k_BT_K$
or the bandwidth of quasiparticles. According to previous
papers,\cite{ohkawaCW,miyai} an almost $T$-linear dependence of
$J_Q(+i0, {\bf q})$ at $T\ll T_K$ in a small region of ${\bf q}$,
${\bf q} \simeq 0$ for ferromagnets and  ${\bf q} \simeq {\bf Q}$ for
antiferromagnets, with ${\bf Q}$ being the nesting wavenumber,  is
responsible for the Curie-Weiss law of itinerant-electron magnets; the
$T$-linear dependence of $1/\tilde{\chi}_s(0)$  at
$T\gg T_K$ is responsible for the Curie-Weiss law of local-moment
magnets. Magnetism with $T_N \gg T_K$ is characterized as local-moment
one, while magnetism with  $T_N\ll T_K$ is characterized as
itinerant-electron one.

%%%%%%%%%%%%%%%%%%%%%%%%%%%%%%%%%%%%%%%%%%%%%%%%%%%%%
%%%%%%%%%%%%%%%%%%%%%%%%%%%%%%%%%%%%%%%%%%%%%%%%%%%%%
%%%%%%%%%%%*%%%%%%%%%%%%%%%%%%%%%%%%%%%%%%%%%%%%%%%%%%
%%%%%%%%%%%%%%%%%%%%%%%%%%%%%%%%%%%%%%%%%%%%%%%%%%%%%

%%%%%%%%%%%%%%%%%%%%%%%%%%%%%%%%%%%%%%%%%%%%%%%%%%%%%
%%%%%%%%%%%%%%%%%%%%%%%%%%%%%%%%%%%%%%%%%%%%%%%%%%%%%
%%%%%%%%%%%*%%%%%%%%%%%%%%%%%%%%%%%%%%%%%%%%%%%%%%%%%%
%%%%%%%%%%%%%%%%%%%%%%%%%%%%%%%%%%%%%%%%%%%%%%%%%%%%%

\subsection{Mode-mode coupling terms}
\label{SecMode}

Following previous papers,\cite{miyai,kawabata,OhkawaModeMode} we consider
mode-mode coupling terms linear  in intersite spin fluctuations
$F_s (i\omega_l, {\bf q}) $ given by Eq.~(\ref{EqF}): 
\begin{equation}
\Lambda (i \omega, {\bf q}) =
\Lambda_L (i \omega_l)
+ \Lambda_s (i \omega_l, {\bf q})
+ \Lambda_v (i \omega_l, {\bf q}) .
\end{equation}
The first term $\Lambda_L (i \omega_l)$ is a {\it local} mode-mode
coupling term, which includes a single {\it local} four-point vertex
function, as is  shown in Fig.~4 of
Ref.~\onlinecite{OhkawaModeMode}.
Both of 
$\Lambda_s (i \omega_l, {\bf q})$ and
$\Lambda_v (i \omega_l, {\bf q})$ are {\it intersite} mode-mode coupling
terms, which include a single {\it intersite} four-point vertex function;
a single $F(i\omega_l,{\bf q})$
appears as the selfenergy correction to the single-particle Green
function in $\Lambda_s (i \omega_l, {\bf q})$
while it appears as a vertex correction to 
the polarization function  in
$\Lambda_v (i \omega_l, {\bf q})$, as are shown in Figs.~3(a) and 3(b),
respectively,  of Ref.~\onlinecite{OhkawaModeMode}.
Their static components are given by
\small 
\begin{equation}\label{Lambda_L}
\Lambda_L(0)
= \frac{5}{2\tilde{\chi}_s(0)}\frac{k_B T}{N}
\sum_{\omega_l{\bf q}} F_s(i\omega_l,{\bf q}) ,
\end{equation}
\begin{eqnarray}\label{Lambda_s}
\Lambda_s(0,{\bf q}) &=&
\frac{3}{\tilde{\chi}_s(0)} 
\frac{k_BT}{N} \sum_{\omega_l, {\bf p}}
B_s(i\omega_l,{\bf p};{\bf q})
\nonumber \\ && \times \left[
F_s (i\omega_l, {\bf q}) 
- \frac1{4}J({\bf q}) \tilde{\chi}_s^2 (i\omega_l)
\right],
\end{eqnarray}
\normalsize
and 
\small
\begin{equation}\label{Lambda_v}
\Lambda_v(0,{\bf q}) =
-\frac{1}{2\tilde{\chi}_s(0)} 
\frac{k_BT}{N} \sum_{\omega_l, {\bf p}}
B_v(i\omega_l,{\bf p};{\bf q})
F_s(i\omega_l, {\bf q}) ,
\end{equation}
\normalsize
with 
\small 
\begin{eqnarray}\label{EqBs}
B_s(i\omega_l,{\bf p};{\bf q}) &=&
\frac{4\tilde{W}_s^4}{\tilde{\chi}_s^3(0)}
k_BT \sum_{\varepsilon_n} \Biggl\{
\frac1{N} \sum_{\bf k}
g_\sigma^{(0)}(i\varepsilon_n,{\bf k}-{\bf q})
\nonumber \\ && \times 
\left[g_\sigma^{(0)}(i\varepsilon_n,{\bf k})\right]^2
g_\sigma^{(0)}(i\varepsilon_n+i\omega_l,{\bf k}+{\bf p})
\nonumber \\ && 
- \left[r_\sigma^{(0)}(i\varepsilon_n)\right]^3
r_\sigma^{(0)}(i\varepsilon_n+i\omega_l)
\Biggr\} ,
\end{eqnarray}
\normalsize
and
\small 
\begin{eqnarray}\label{EqBv}
B_v(i\omega_l,{\bf p};{\bf q}) &=&
\frac{4\tilde{W}_s^4}{\tilde{\chi}_s^3(0)}
k_BT \sum_{\varepsilon_n} \Biggl\{
\frac1{N} \sum_{\bf k}
g_\sigma^{(0)}(i\varepsilon_n,{\bf k}+{\bf q})
\nonumber \\  && \times 
g_\sigma^{(0)}(i\varepsilon_n,{\bf k})
g_\sigma^{(0)}(i\varepsilon_n+i\omega_l,{\bf k}+{\bf q}+{\bf p})
\nonumber \\  && \times 
g_\sigma^{(0)}(i\varepsilon_n+i\omega_l,{\bf k}+{\bf p})
\nonumber \\ && 
- \left[r_\sigma^{(0)}(i\varepsilon_n)\right]^2 \!
\left[r_\sigma^{(0)}(i\varepsilon_n \!+\! i\omega_l)\right]^2
\Biggr\} ,
\end{eqnarray}
\normalsize 
with
\begin{equation}
r_\sigma^{(0)}(i\varepsilon_n) = 
\frac1{N} \sum_{\bf k} g_\sigma^{(0)}(i\varepsilon_n, {\bf k}).
\end{equation}
Because the selfenergy
correction linear in $J({\bf q})$ is considered in
Sec.~\ref{SecFock}, 
$\frac1{4}J({\bf q})\tilde{\chi}_s^2(i\omega_l)$ is subtracted
in Eq.~(\ref{Lambda_s}).

In this paper, weak three dimensionality  in spin fluctuations is
phenomenologically included.  Because
$J({\bf q})$ has its maximum value at 
${\bf q}= (\pm \pi/a, \pm\pi/a)$ and the nesting vector of the
Fermi surface in two
dimensions is also close to 
${\bf q}=(\pm \pi/a, \pm \pi/a)$ for almost half filling, we assume
that the ordering wave number in three dimensions is 
\begin{equation}\label{EqIncommensurateQ} 
{\bf Q} = (\pm \pi/a, \pm \pi/a, \pm Q_z), 
\end{equation}
with $Q_z$ depending on interlayer exchange interactions. 
On the phase boundary between paramagnetic and antiferromagnetic
phases, where Eq.~(\ref{EqAFCondition}) is satisfied, the
inverse of the susceptibility is expanded around ${\bf Q}$ and
for small $\omega_l$ in such a way that
\begin{equation}
\left[1/\chi_s(i\omega_l,{\bf Q} + {\bf q})\right]_{T=T_N} =
A({\bf q}) + \alpha_\omega |\omega_l|  + \cdots,
\end{equation}
with
\begin{equation}\label{EqAQ}
A({\bf q}) = \frac1{4}A_\parallel ({\bf q}_\parallel a)^2 
+ \frac1{4} A_z \left[( q_z - Q_z) c\right]^2 .
\end{equation}
Here, $c$ is the lattice constant along the $z$ axis. 
Because $\chi_s(i\omega_l,{\bf Q} + {\bf q})$ diverges in
the limit of $|{\bf q}| \rightarrow 0$ and $\omega_l \rightarrow 0$
on the phase boundary, 
\small
\begin{eqnarray}
B_s(0,-{\bf Q};{\bf Q}) &=&
B_v(0,-{\bf Q};{\bf Q})
\nonumber \\ &=& 
\frac{4\tilde{W}_s^4}{\tilde{\chi}_s^3}
k_B T \sum_{\varepsilon_n} \Biggl\{
\frac1{N}\! \sum_{\bf k}
\left[g_\sigma^{(0)}(i\varepsilon_n,{\bf k}\!-\!{\bf Q})\right]^2
\nonumber \\ && \times 
\left[g_\sigma^{(0)}(i\varepsilon_n,{\bf k}) \right]^2
-\left[r_\sigma^{(0)} (i\varepsilon_n)\right]^4
\Biggr\} ,
\end{eqnarray}
\normalsize
can be approximately used for $B_s(i\omega_l,{\bf p};{\bf q})$ in
Eq.~(\ref{Lambda_s}) and 
$B_v(i\omega_l,{\bf p};{\bf q})$ in Eq.~(\ref{Lambda_v}).   Then,  it
follows that
\begin{equation}
\Lambda (0,{\bf Q}) =
\frac{5}{2\tilde{\chi}_s(0)} (1 + C_F - \tilde{C} _L) \Phi ,
\end{equation} 
with
\begin{eqnarray}\label{EqCF}
C_F &=& 
\frac{8W_s^4}{\tilde{\chi}_s^3(0)}
\frac{1}{N} \sum_{\bf k}
\Biggl\{
\frac{f_{\gamma}(\xi({\bf k} \!+\! {\bf Q}) \!-\! \mu^*)
\!-\! f_{\gamma}(\xi({\bf k}) \!-\! \mu^*)}
{\xi({\bf k})-\xi({\bf k}+{\bf Q})}
\nonumber\\ && \quad  +
\frac1{2}
\Bigl[ f_{\gamma}^\prime(\xi({\bf k}) \!-\! \mu^*)
+f_{\gamma}^\prime(\xi({\bf k} \!+\! {\bf Q}) \!-\! \mu^*)\Bigr] \!
\Biggr\}
\nonumber\\ && \qquad \times 
\frac1{[\xi({\bf k})-\xi({\bf k}+{\bf Q})]^2} ,
\end{eqnarray}
\begin{equation}\label{EqCL}
\tilde{C}_L = \frac{16 \tilde{W}_s^4}{\tilde{\chi}_s^3(0)}
\!\int \!\!d\varepsilon \Bigl[
\bar{\rho}(x)\bar{\rho}_2^3(\varepsilon) -
\pi^2\bar{\rho}^3(\varepsilon)\bar{\rho}_2(\varepsilon)
\Bigr] f_\gamma(\varepsilon-\mu^*) ,
\end{equation}
and
\begin{eqnarray}\label{EqPhi}
\Phi &\equiv&  
%\frac{k_BT}{N} \sum_{\omega_l{\bf q}} F_s (i\omega_l,{\bf q}) 
%\simeq 
\frac{k_BT}{N} \sum_{\omega_l}~\!^{\Large\prime} 
\sum_{|{\bf q}|\le q_c} \sum_{q_z}
\frac1{A({\bf q}) + \alpha_\omega |\omega_l|}
\nonumber \\ &=&
\frac{2 c}{\pi^3 A_\parallel} \hspace{-2pt}
\int_0^{\pi/c} \hspace{-5pt} dq_z 
\int_0^{\omega_c} \hspace{-5pt} d\omega 
\left[ n(\omega) +\frac1{2} \right] \!
\nonumber \\ && \times 
\Biggl\{\tan^{-1}\!
\left[ \frac{A_\parallel (q_c a)^2 + A_z (q_zc)^2}
{4\alpha_\omega\omega}\right] 
\nonumber \\ && \qquad 
- \tan^{-1}\!\left[ \frac{A_z (q_zc)^2}
{4\alpha_\omega\omega}\right]
\Biggr\} .
\end{eqnarray}
In Eq.~(\ref{EqCF}), $f_{\gamma}^\prime (\varepsilon)$  is the
derivative of $f_\gamma(\varepsilon)$ defined by Eq.~(\ref{EqfGamma}):
\begin{equation}
f_{\gamma}^\prime (\varepsilon) = 
-\frac1{2\pi^2 k_B T} \mbox{Re}  \left[
\psi^{\prime} \left( \frac1{2} 
+ \frac{\gamma  -i \varepsilon }{2\pi k_B T} \right)
\right] .
\end{equation}
In Eq.~(\ref{EqCL}), $\rho_{\gamma\rightarrow 0}(\varepsilon)$ is simply
denoted by $\bar{\rho}(\varepsilon)$, and 
\begin{equation}
\bar{\rho}_2(\varepsilon) = \mbox{Vp}\! \int d\varepsilon^\prime 
\frac{\bar{\rho}(\varepsilon^\prime)}
{\varepsilon-\varepsilon^\prime} .
\end{equation}
In Eq.~(\ref{EqPhi}), the summation is restricted to
$|\omega| \le \omega_c$ and 
$|{\bf q}_\parallel| \le q_c $. 
We assume that $q_c=\pi/3a$ and $\omega_c$ is given by a larger one of
$8[t^*|$ and  $|J|$.

As was shown in the previous paper,\cite{OhkawaPseudogap}  
$\alpha_\omega\simeq 1$ for $T/T_K \alt 1$ and $\gamma/k_BT_K \alt 1$.
Physical properties for $T/T_K \gg 1$ or $\gamma/k_BT_K \gg 1$
scarcely depends on $\alpha_\omega$. Then, we assume  $\alpha_\omega
=1$ for any $T$ and $\gamma$ in this paper. 

It is easy to confirm that $\Phi$ diverges at non zero temperatures
for $A_z=0$.  No magnetic instability occurs at nonzero temperature in
two dimensions because of the divergence of the mode-mode coupling
term.
 
When only the ${\bf q}$ dependence of the superexchange interaction  is
considered,  $A_\parallel = |J|$.  Because  $(1/4)J_Q(0,{\bf Q}+{\bf
q})$ also contribute to the $q$-quadratic term, $A_\parallel$ is
larger than $|J|$. However, we assume $A_\parallel = |J|$ for the sake
of simplicity and
%%%%%%%%%%%%%%%%%%%%%%%%%%%%%%%%%%%%%%%%%%%%%%%%%%%%%%%%%%%%%%
\begin{equation}\label{Eq2D-SpinFluct}
|A_z/A_\parallel| = 10^{-10} 
\end{equation}
in order to reproduce observed $T_N$ for the just half filling: 
$k_BT_N/|t| \simeq 0.06\simeq 0.2 |J|/|t| $.

%%%%%%%%%%%%%%%%%%%%%%%%%%%%%%%%%%%%%%%%%%%%%%%%%%%%%
%%%%%%%%%%%%%%%%%%%%%%%%%%%%%%%%%%%%%%%%%%%%%%%%%%%%%
\begin{figure} 
\centerline{\hspace*{1.2cm}
\includegraphics[width=7.5cm]{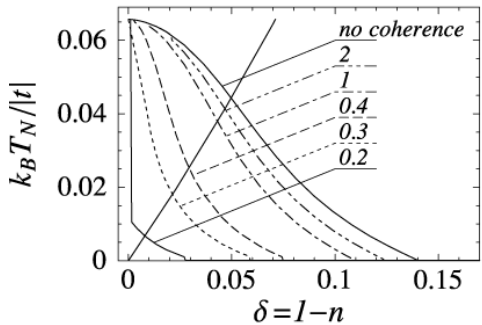}
}
\caption[6]{ 
$T_N$ of the symmetric model as a function of $\delta= 1-n$.  From the
bottom, solid, dotted, broken, dot-broken, and double-dot-broken lines
show $T_N$ for $\gamma/|t|=0.2$, 0.3, 0.4, 1, and 2, respectively. The
topmost solid line shows $T_N$ determined from
Eq.~(\ref{EqAFCondition2}) for comparison.  The solid line with a
positive slope shows $1/\tilde{\phi}_\gamma$.  Antiferromagnetic states
whose $T_N$ are much below this line are, at least, characterized as
itinerant-electron ones. }
\label{T_N}
\end{figure} 
%%%%%%%%%%%%%%%%%%%%%%%%%%%%%%%%%%%%%%%%%%%%%%%%%%%%%
%%%%%%%%%%%%%%%%%%%%%%%%%%%%%%%%%%%%%%%%%%%%%%%%%%%%%
\begin{figure} 
\centerline{\hspace*{1cm}
\includegraphics[width=7.5cm]{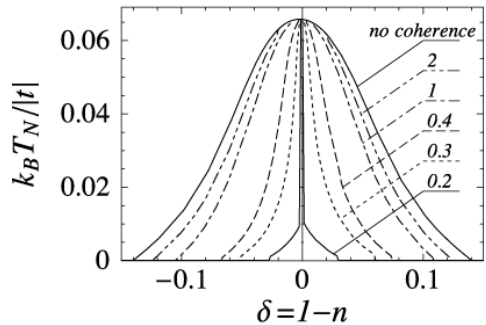}
}
\caption[7]{ 
$T_N$ of the asymmetric model as a function of $\delta$.  See also the
caption of Fig.~\ref{T_N}. Note that $T_N$ is almost symmetric between
$\delta<0$ and $\delta>0$.

 }
\label{T_N-asym}
\end{figure} 
%%%%%%%%%%%*%%%%%%%%%%%%%%%%%%%%%%%%%%%%%%%%%%%%%%%%%%
%%%%%%%%%%%%%%%%%%%%%%%%%%%%%%%%%%%%%%%%%%%%%%%%%%%%%

\subsection{Almost symmetric $T_N$ between $\delta>0$ and $\delta<0$}

When the coherent part of
the Green function is not considered,  $\tilde{C}_L=0$ and
$C_F=0$; the instability condition
(\ref{EqAFCondition}) becomes as simple as
\begin{equation}\label{EqAFCondition2}
\frac1{\tilde{\chi}_s(0)} + \frac{5\Phi}{2\tilde{\chi}_s(0)} 
- \frac1{4}J({\bf Q}) =0 .
\end{equation}
When the mode-mode coupling term $5\Phi/2\tilde{\chi}_s(0)$ is ignored and
we assume $T_K =0$,  Eq.~(\ref{EqAFCondition2}) gives $T_N = (1/4)J({\bf
Q})/k_B=|J|/k_B$, which is nothing but $T_N$ in the mean-field
approximation for the Heisenberg model. Because Eq.~(\ref{EqAFCondition2})
is what is expected for 
$T/T_K\gg 1$, Eq.~(\ref{EqAFCondition}) is
valid for not only $T \ll T_K$ but also $T \gg T_K$. Then, 
it is, at least, qualitatively valid even for the crossover
region between $T\ll T_K$ and $T\gg T_K$.

It is possible that $\gamma/|t| = O(1)$, as is examined in
Appendix~\ref{SecDisorder}.  Figure~\ref{T_N} shows $T_N$ of the symmetric
model as a function of 
$\delta=1-n$ for various $\gamma$.  The region of
antiferromagnetic states is wider for larger $\gamma$, mainly because $T_K$
is lower for larger $\gamma$. Figure~\ref{T_N-asym} shows $T_N$ of the
asymmetric model as a function of $\delta$. As long as $\gamma$ is almost
symmetric with respect to
$\delta$, $T_N$ is also almost symmetric with respect to $\delta$. The
difference of the Fermi surfaces cannot give any significant 
asymmetry of $T_N$ between $\delta <0$ and $\delta>0$.

\section{Application to cuprate oxides}
\label{SecDiscussion}

The $t$-$J$ model with the just half filling is reduced to the
Heisenberg model.  The N\'{e}el temperature is as
high as $T_N = |J|/k_B$ in the mean-field approximation for the
Heisenberg model. It is much higher than observed $T_N$. This
discrepancy can be explained by the reduction of $T_N$ by quasi-two
dimensional thermal spin fluctuations or by the local mode-mode
coupling term $\Lambda_L(0)$. When the anisotropy is as large as 
Eq.~(\ref{Eq2D-SpinFluct}),  $T_N$ is as low as  $T_N\simeq 0.2
|J|/k_B$, as is shown in Fig.~\ref{T_N}.
This explains observed $T_N \simeq 300$~K, when we take $|J|\simeq
0.15$~eV. However, the assumed anisotropy  seems to be a little too
large.  We should consider the reduction of $T_N$ more properly than
we do in this paper.

When electrons or holes are doped, Gutzwiller's quasiparticles are formed
on the chemical potential. When their life-time widths $\gamma$ or
temperatures $k_BT$ are much larger than their  bandwidth, however,
quasiparticles can play no significant role. The Kondo temperature is
approximately given by $k_BT_K \simeq 2|\delta t|$,
with $\delta$ being the concentrations of dopants. The N\'{e}el
temperature $T_N$ is determined by the competition between the
stabilization of antiferromagnetism by the superexchange interaction $J$
and the quenching of magnetic moments by the Kondo effect with $k_BT_K$
and the local mode-mode coupling term $\Lambda_L(0)$.  The reduction of
$T_N$ for small $|\delta|$ is mainly due to $\Lambda_L(0)$, as is
discussed above. On the other hand, the critical concentration
$|\delta_c|$ below which antiferromagnetic ordering appears at $T=0$~K is
mainly determined by the competition between $J$ and $k_B T_K$, because
thermal spin fluctuations vanish at $T=0$~K.  The critical concentration 
is as large as $|\delta_c|\simeq 0.14$  for parameters relevant for
cuprates, as is shown in Figs.~\ref{T_N} and \ref{T_N-asym}.

When both of $\gamma$ and $k_BT$ are small, the selfenergy-type
mode-mode coupling term $\Lambda_s(0,{\bf Q})$ is large. Not only the
linear terms in  
$F_s(i\omega_l,{\bf q})=\chi_s(i\omega_l,{\bf q})-\tilde{\chi}_s(0)$  
but also higher order terms in $F_s(i\omega_l,{\bf q})$ should be
considered, for example, in a FLEX approximation. The N\'{e}el
temperature $T_N$ is a little higher in the FLEX approximation than it
is in the treatment of this paper.  Unless both of $\gamma$ and $k_BT$
are very small, $\Lambda_s(0,{\bf Q})$ is much smaller than
$\Lambda_L(0)$.  In such a case, no significant correction can arise
even if the treatment of $\Lambda_s(0,{\bf Q})$ itself is irrelevant. 
Relative corrections $\Delta T_N/T_N$ are large only
for low $T_N$, for example $T_N/|t|\alt 0.005$, with $\Delta T_N$ an
increment of $T_N$ in the FLEX approximation. Because $\Delta T_N/T_N$
are large only in such low $T_N$ regions, corrections  $\Delta T_N$
themselves are never significant. Note that $T_N$ for large enough $\gamma$ or $T_N$ determined from
Eq.~(\ref{EqAFCondition2}) has no correction in the FLEX
approximation.  

When $\gamma$ and $k_BT$ are much smaller than the quasiparticle
bandwidth, quasiparticles can play significant roles in not only the
enhancement but also the suppression of antiferromagnetism.
Antiferromagnetism is enhanced by the exchange interaction 
$J_Q(i \omega_l, {\bf q})$ airing from the virtual exchange of pair
excitations of quasiparticles, in which the nesting of the Fermi
surface can play a significant role. The Fock term of the
superexchange interaction renormalize quasiparticles;
the bandwidth of quasiparticles is approximately given by
$8|t^*|$, with
\begin{equation}\label{EqBandwidth2} 
t^*  \simeq  (\pi^2/8)|\delta t| -
\left(3\tilde{W}_s^2/2\pi^2\right) J ,
\end{equation}
in the limit of $\gamma/|t^*|\rightarrow +0$ and $k_BT/|t^*|\rightarrow
+0$. The Kondo temperature is approximately  given 
by $k_BT_K\simeq (1/4c_{T_K})|t^*|\simeq 1.6|t^*|$. The Kondo effect of
quenching magnetism is stronger when $\gamma$ are smaller or
quasiparticles are more itinerant. The intersite mode-mode coupling term,
$\Lambda_s(0,{\bf Q})+\Lambda_v(0,{\bf Q})$,
also suppresses antiferromagnetism in addition to the local mode-mode
coupling term $\Lambda_L(0)$. Because the quenching effects overcome the
enhancement effect, $T_N$ decrease with decreasing $\gamma$.

Physical properties  are asymmetric between electron-doped 
$(\delta<0)$ and hole-doped  $(\delta>0)$ cuprates.  For example, an
antiferromagnetic states appears in a narrow range of
$0\le |\delta|\alt 0.02$--$0.05$ in hole-doped cuprates,  while it
appears in a wide range of $0\le|\delta| \alt 0.13$--$0.15$ in
electron-doped cuprates. Tohyama and Maekawa\cite{tohyama} argued that
the asymmetry must arises from the difference of the Fermi surfaces,
and that  the $t$-$t^\prime$-$J$ or asymmetric model should be used.
They showed that  the intensity of spin excitations is relatively
stronger in electron doping cases than it is in hole doping cases. 
According to Fig.~\ref{pol-asy}, the polarization function at
${\bf q}=(\pm\pi/2a,\pm\pi/2a)$  is relatively larger in electron
doping cases than it is in hole doping cases. This asymmetry is
consistent with that of spin excitations studied by Tohyama and
Maekawa.   However, the difference of
the Fermi surfaces cannot explain the asymmetry of $T_N$, as
is shown in Fig.~\ref{T_N-asym}.

The condensation energy at $T=0$~K of the asymmetric model is
also quite asymmetric;\cite{yokoyama} it is consistent with the asymmetry
discussed above. On the other hand,  
$T_N$ is significantly reduced by quasi-two dimensional spin fluctuations
as well as local spin fluctuations of the Kondo effect. This large
reduction of $T_N$ arises from the renormalization of normal states; not
only the N\'{e}el states but also paramagnetic states just above
$T_N$ are largely renormalized by the spin fluctuations.  It is
plausible that the asymmetry of the condensation energy of paramagnetic
states just above
$T_N$  is similar to that of the N\'{e}el states at $T=0$~K. It is
interesting to  confirm by comparing the condensation energy
of the N\'{e}el states and that of paramagnetic states just above $T_N$
whether $T_N$ is actually almost symmetric as is shown in this paper.

Electrical resistivities of electron-doped cuprates are relatively
larger than those of hole-doped cuprates are.\cite{tokura}  A
plausible explanation is that the asymmetry of $T_N$  arises mainly
from the difference of disorder.  Critical $|\delta|$'s for
electron-doped cuprates below which  an antiferromagnetic state appears
are as large as $0.13$--$0.15$. These numbers
are close to the theoretical critical value about
$|\delta_c|=0.14$ for large $\gamma$. This implies that disorder of
electron-doped cuprates must be large.   In hole-doped cuprates, on
the other hand, an antiferromagnetic state appear only in a narrow
range of $0\le\delta\alt 0.02$--0.05 although antiferromagnetic spin
fluctuations are well developed in a wide range of $0\le\delta\alt
0.15$. This implies that a paramagnetic state in the range of
$0\le\delta\alt 0.15$ is in the vicinity of an antiferromagnetic
critical point. We expect that when disorder is introduced into such
paramagnetic hole-doped cuprates  antiferromagnetism must appear in a
wide range of hole concentrations. In actual, magnetic moments appear
when Zn ions are introduced.\cite{CuprateZn}   An almost symmetric
behavior of $T_N$ must be restored by preparing hole-doped and
electron-doped cuprates with similar degree of disorder to each other.

\begin{figure} 
\centerline{\hspace*{0.1cm}
\includegraphics[width=4.9cm]{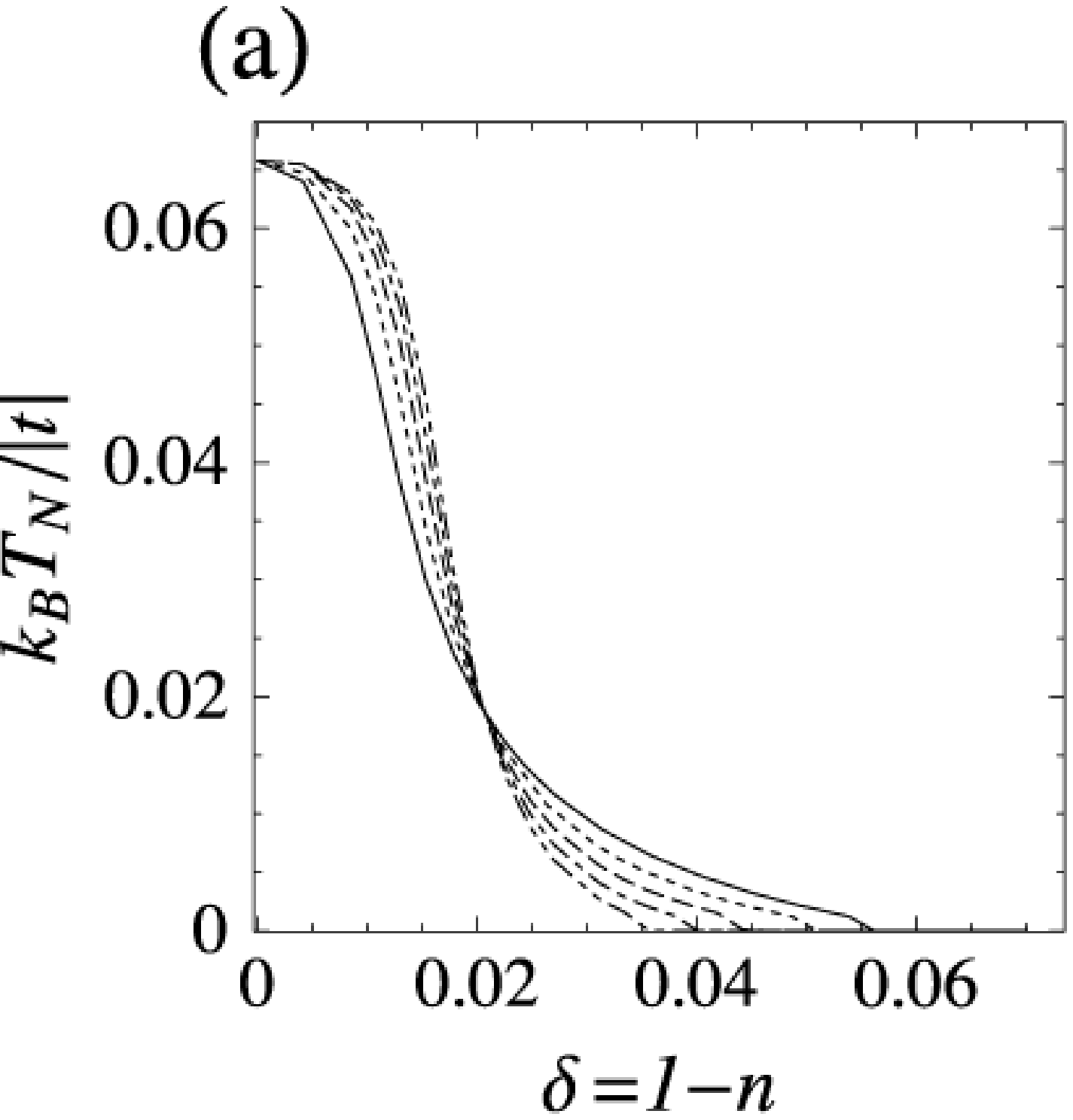}
\hspace*{-0.7cm}
\includegraphics[width=4.8cm]{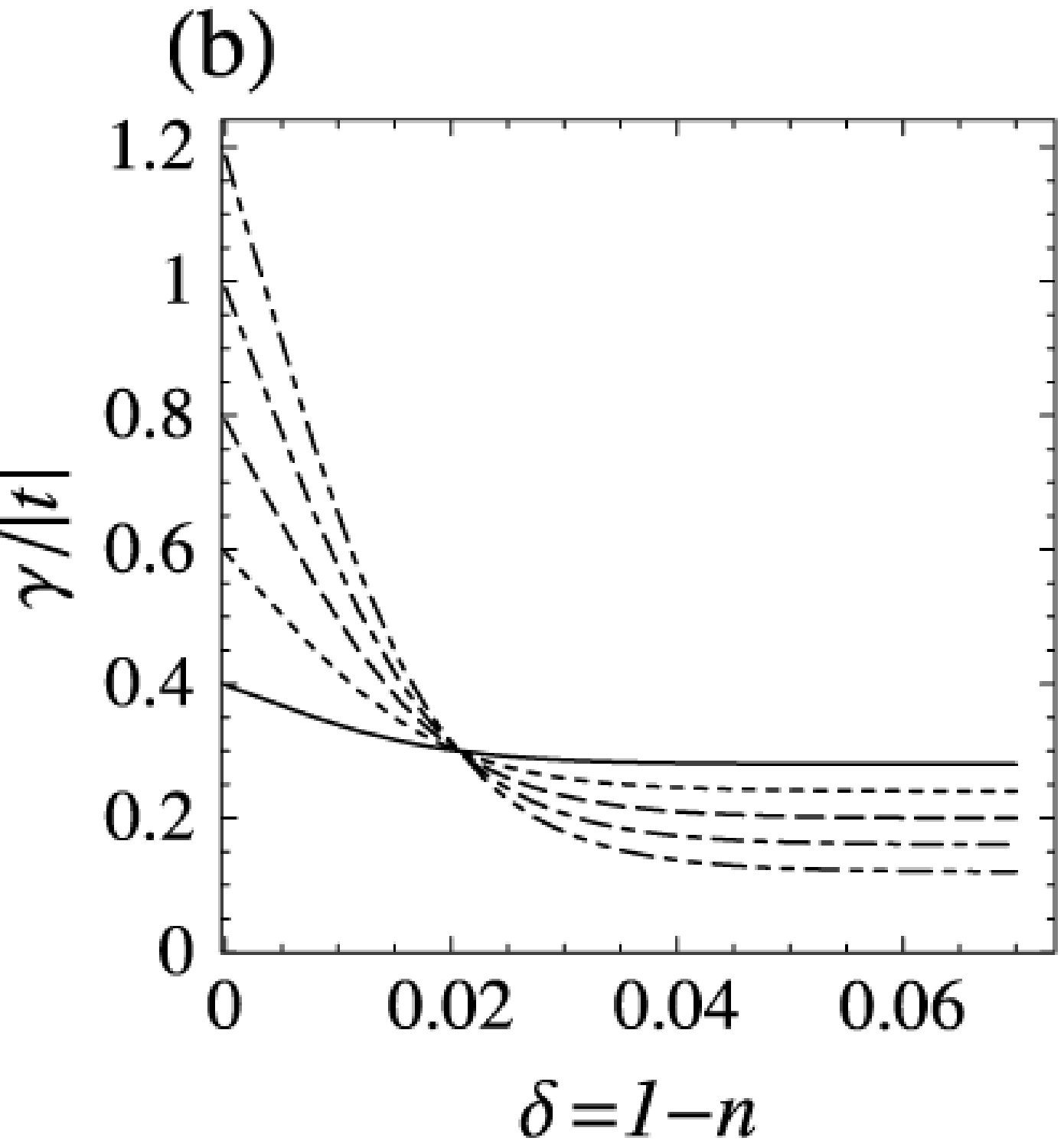}
\hspace*{-0.8cm}
}
\caption[8]{ 
(a) $T_N$ as a function of $\delta$, and 
(b) $\gamma$ as a function of $\delta$.
When we assume $\gamma$ shown in Fig.~\ref{Kaumagai's}(b), we obtain
$T_N$ shown  in Fig.~\ref{Kaumagai's}(a)  by the same kind of line as that
for $\gamma$.
 }
\label{Kaumagai's}
\end{figure}

An antiferromagnetic state in the range of $0\le\delta\alt
0.02$ of hole-doped La$_{2-\delta}$M$_\delta$CuO$_4$
(M= Sr or Ba) is characterized as a local-moment one. The so called
spin-glass or Kumagai's phase\cite{Kumagai} appears in the range of
$0.02\alt\delta\alt 0.05$. The $\delta$ dependence of
$T_N$ observed for hole-doped cuprate is qualitatively different from
that of theoretical $T_N$ shown in Figs.~\ref{T_N} and \ref{T_N-asym},
where $\gamma$ is assumed to be constant as a function of $\delta$.
Experimentally, electrical resistivities are large for under-doped
cuprates. This observation implies that in general $\gamma$ is a
decreasing function of $|\delta|$. For example,
Fig.~\ref{Kaumagai's}(a) shows theoretical $T_N$ for several cases of
$\gamma(\delta)$ as a function of $\delta$, which are shown in
Fig.~\ref{Kaumagai's}(b). If we take a proper $\gamma(\delta)$,
observed $T_N(\delta)$, including
$T_N(\delta)$ of Kumagai's phase, can be reproduced.

%%%%%%%%%%%*%%%%%%%%%%%%%%%%%%%%%%%%%%%%%%%%%%%%%%%%%%
%%%%%%%%%%%%%%%%%%%%%%%%%%%%%%%%%%%%%%%%%%%%%%%%%%%%%
%%%%%%%%%%%*%%%%%%%%%%%%%%%%%%%%%%%%%%%%%%%%%%%%%%%%%%
%%%%%%%%%%%%%%%%%%%%%%%%%%%%%%%%%%%%%%%%%%%%%%%%%%%%%
\begin{figure} 
\centerline{\hspace*{1.2cm}
\includegraphics[width=7.5cm]{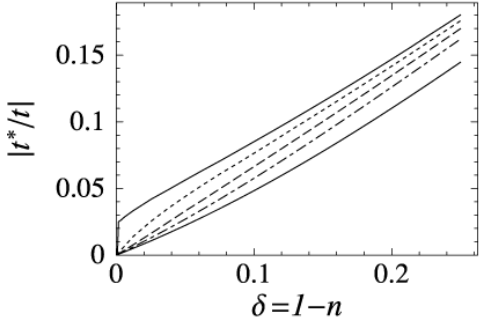}
}
\caption[9]{ 
$t^*$ of the symmetric model as a function of $\delta$ for various $\gamma$
and $k_BT/|t|=0.02$; $\tilde{W}_s\simeq 1$ or $r=0.5$ is assumed instead of
$\tilde{W}_s\simeq 2$ or  $r=1$ (See text). From the top, solid, dotted,
broken, and dot-broken line show $t^*$ for
$\gamma/|t|=0.04$,  0.1, 0.2, and 0.4, respectively.  For the sake of
comparison, $1/\tilde{\phi}_\gamma$ is also shown by a bottom solid line. 
}
\label{tstar-1}
\end{figure} 
%%%%%%%%%%%%%%%%%%%%%%%%%%%%%%%%%%%%%%%%%%%%%%%%%%%%%
%%%%%%%%%%%%%%%%%%%%%%%%%%%%%%%%%%%%%%%%%%%%%%%%%%%%%
\begin{figure} 
\centerline{\hspace*{1.2cm}
\includegraphics[width=7.5cm]{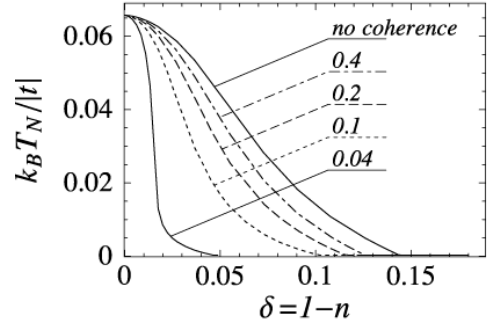}
}
\caption[9]{ 
$T_N$ of the symmetric model as a function of $\delta$ for various
$\gamma$; $\tilde{W}_s\simeq 1$ or  $r=0.5$ is assumed instead of
$\tilde{W}_s\simeq 2$ or $r=1$ (See text). From the bottom, solid,
dotted, broken, and dot-broken lines show $T_N$ for $\gamma/|t|=0.04$,
0.1, 0.2, and 0.4, respectively. For comparison,  $T_N$
determined from Eq.~(\ref{EqAFCondition2}) is shown by a topmost
solid line.     }
\label{T_N3}
\end{figure} 
%%%%%%%%%%%%%%%%%%%%%%%%%%%%%%%%%%%%%%%%%%%%%%%%%%%%%
%%%%%%%%%%%%%%%%%%%%%%%%%%%%%%%%%%%%%%%%%%%%%%%%%%%%%

The effective three-point vertex function $\tilde{\phi}_s$ and the
mass-renormalization factor $\tilde{\phi}_\gamma$ are those in SSA. They
are renormalized by intersite fluctuations such as antiferromagnetic and
superconducting fluctuations.  We can take into account these intersite types of
renormalization phenomenologically,  or we can treat $\tilde{W}_s$ as a phenomenological
parameter following the previous paper,\cite{OhkawaPseudogap} where
$\tilde{W}_s=0.7\mbox{--}1$ is used in stead of $\tilde{W}_s\simeq 2$ in
order to explain quantitatively observed superconducting critical
temperatures and $T$-linear resistivities of cuprates.  
Then, we replace $\tilde{W}_s$ by
$r\tilde{W}_s$, with $r$ a numerical constant smaller
than unity;\cite{comment-Ws} Eq.~(\ref{Eqt*}) is replaced by
\begin{equation}\label{EqReplace}
2 t^* = r \frac{2t}{\tilde{\phi}_\gamma} 
- \frac{3}{4}\left(r\tilde{W}_s\right)^2 J \Xi. 
\end{equation}
%%%%%%%%%%%%%%%%%%%%%%%%%%%%%%%%%%%%%%%%%%%%%%%%%%%%%%%%%%%%%%%%%%%%%
Figures~\ref{tstar-1} and \ref{T_N3} show $t^*$ and $T_N$, respectively, 
of the symmetric model as a function of $\delta$ for $r=0.5$ and various
$\gamma$.  
The antiferromagnetic region extends with decreasing $r$, but
theoretical curves for $r=0.5$ shown in  Fig.~\ref{T_N3} are qualitatively
the same as those for $r=1$ shown in Fig.~\ref{T_N}.   When we take a
proper $\gamma(\delta)$, we can also reproduce observed $T_N(\delta)$ even
for $r=0.5$ or $\tilde{W}_s\simeq 1$. 

According to Fig.~\ref{T_N3}, $\gamma/|t|\simeq 0.04$ is needed in order
to reproduce Kumagai's phase. According to Fig.~\ref{tstar-1},
$|t^*/t|\simeq 0.05$ for $\delta=0.04$. Then, we can argue
that  $k_F l \simeq 2k_BT_K/\gamma \simeq 4|t^*|/\gamma$, with $k_F$ the
Fermi wave number and
$l$ the mean free path, must be 4--8 in Kumagai's
phase.\cite{ComGammaE-dep} According to Fig.~\ref{pol-sym}(b), the
nesting of the Fermi surface is  substantial at least for $\gamma/|t^*|
\alt 0.3$; the nesting cannot be ignored for $\gamma/|t^*| \alt 1$.
Kumagai's phase must be a spin density wave (SDW) state in a disordered
system rather than a spin glass. The divergence\cite{Kumagai} of the
nuclear quadrupole relaxation (NQR) rate at $T_N$  supports this
characterization.

The so called stripe phase appears in the vicinity of
$\delta =1/8$.\cite{tranquada} Because superconductivity is
suppressed, the pair breaking by disorder must be large;
disorder may be related with a structural phase transition of first
order between high-temperature tetragonal (HTT) and low-temperature
orthorhombic (LTO) lattices.\cite{fujita,fleming} Then, resistivities
must also be large.  In actual, resistivities increase logarithmically
with decreasing temperatures in LTO phase;\cite{uchida} the
logarithmic increase implies the Anderson localization, as is
discussed below.  One of possible explanations is that a SDW state 
enhanced by disorder is stabilized.  According to Fig.~\ref{T_N3},
$\gamma/|t|\simeq 0.2$ is needed in order that an antiferromagnetic
state might appear for $\delta\simeq 1/8$. On the other hand,
$|t^*/t|\simeq 0.1$  for $\delta\simeq 1/8$. Then, we can argue that
when  $k_F l \alt 2$ is satisfied SDW can appear for $\delta\simeq
1/8$.\cite{ComGammaE-dep} Because the strength of quenching of magnetic
moments, $k_BT_K$, is an increasing function of
$|\delta|$, a charge density wave (CDW)  appears in such a way that the
electron filling is closer to unity  at sites where magnetizations are
larger; the wave number of CDW is twice of that of
SDW.\cite{antiferromagnetism} 

Because effective disorder increases with increasing magnetic fields in
disordered Kondo lattices,\cite{OhkawaPositiveMR} where the distribution
or disorderness of $T_K$ is large, we argue that  disordered cuprates
must exhibit large positive magnetoresistance  and antiferromagnetic
ordering must be induced by magnetic fields.   It is 
interesting to examine whether the stripe state exhibits large positive
magnetoresistance  and the critical temperature of the stripe state is
enhanced by magnetic fields.

For the sake of simplicity, we assume that ordering wave numbers {\bf Q}
are commensurate. However,  
{\bf Q} are incommensurate for large $|\delta|$, as is shown in
Fig.~\ref{pol-asy}. When incommensurate {\bf Q} are considered, $T_N$
become a little higher than they are in this paper. According to
Ref.~\onlinecite{antiferromagnetism},  when  {\bf Q} are incommensurate
and a tetragonal lattice distortion is small enough, a double-{\bf Q}
SDW with magnetizations of different {\bf Q} being orthogonal to each
other can be stabilized on each CuO$_2$ plane. It is interesting to
look for such non-colinear magnetic structure.

The Fock term of the superexchange interaction between nearest neighbors
renormalizes only nearest-neighbor $t^*$, but it does not renormalize
next-nearest-neighbor $t_2^*$. The ratio of $t_2^*/t^*$, which is
assumed to be constant in this paper, must depend on various parameters
such as $\gamma$, $T$, $\delta=1-n$, and so on. For
example, Fig.~\ref{tstar-1} shows
$t^*$ as a function of $\delta$ for various $\gamma$, 
$k_BT/|t|=0.02$, and $\tilde{W}_s\simeq 1$ $(r=0.5)$. In case of
$\tilde{W}_s=0.7\mbox{--}1$, the magnitude of the renormalization of $t^*$
due to the Fock term is about a eighth or a fourth of those shown in
Figs.~\ref{t^star} and \ref{t^star-asym}, where $\tilde{W}_s\simeq 2$ is
assumed.  Although the renormalization is
expected to be rather small in actual cuprates, it is interesting to
examine the dependence of the shape of the Fermi surface or line on such
parameters, in particular, on
$\gamma$. The shape of the Fermi surface must be more similar to what the
symmetric model predicts in better samples with smaller
$\gamma$ than it is in worse samples with large $\gamma$.
This argument also leads to a prediction that because of the Fock term, if
the broadening due to $\gamma$ is corrected,  the bandwidth of
quasiparticles must be  larger in better samples than it is in worse
samples.

The bandwidth of Gutzwiller's quasiparticles is of the order of $|J|$
at $T=0$~K for the Hubbard model with no disorder and the just half
filling, even when
$\tilde{\phi}_\gamma\rightarrow+\infty$.\cite{exchByOhkawa} Because
$J=-4t^2/U$ for large enough $U/|t|$, the bandwidth
vanishes only in the limit of $U/|t|\rightarrow +\infty$. We speculate
that, if the Hilbert space is restricted within paramagnetic states,
the disappearance of Gutzwiller's band or the metal-insulator
transition considered by Brinkman and Rice\cite{Brinkman} must occur
at  $U\rightarrow +\infty$. When the critical $U_c$ is infinitely
large, no hidden order parameter is required in this transition of
{\it second} order. When small disorder is introduced or temperatures
are slightly raised, the transition at $U\rightarrow +\infty$ must
turn out to a crossover around a finite $U$; the $U$  must be very
close to what Brinkman and Rice's theory predicts, unless disorder and
$k_BT$ are extremely small.

In actual metal-insulator transitions, which are of first order, the
symmetry of a lattice changes or the lattice parameter discontinuously
changes.  Within a single-band model with no electron-phonon interaction,
it is difficult, presumably impossible, to reproduce first-order
metal-insulator transitions.  The electron-phonon interaction as well as
orbital degeneracy should be included in order to explain actual
metal-insulator transitions of first order.

One of the most serious assumptions in this paper is the homogeneous
life-time width of quasiparticles. This assumption is irrelevant when
the concentration of dopants is small.  For example, introduce a
single metal ion such as Sr in a purely periodic La$_2$CuO$_4$. A hole
must be bound around the metal ion. When the concentration of metal
ions is small enough, each hole must be similarly bound around each
metal ion or it is in one of the Anderson localized states.  The
Anderson localization may play a role at low temperatures; we expect
the logarithmic dependence of  resistivities on temperatures in two
dimensions as well as negative magnetoresistance.  In actual,
logarithmic divergent resistivities with decreasing temperatures are
observed.\cite{ono} However, observed magnetoresistance is positive
rather than negative.\cite{miura}  Because  positive magnetoresistance
is expected in disordered Kondo lattices,\cite{OhkawaPositiveMR} as is
discussed above, one of the
possible explanations is that the positive magnetoresistance due to
the disorderness of $T_K$  cancels or overcomes the negative
magnetoresistance due to the Anderson localization. The Anderson
localization in disordered Kondo lattices should be seriously
considered to clarify electronic properties of cuprates with rather
small concentrations of dopants or under-doped cuprates.

\section{Conclusion}
\label{SecConclusion}

Effects of life-time widths $\gamma$ of quasiparticles  on 
the N\'{e}el temperature $T_N$ of the
$t$-$J$ model with $J/|t|=-0.3$ on a quasi-two dimensional lattice are
studied within a theoretical framework of Kondo lattices.  The Kondo
temperature $T_K$, which is a measure of  the strength of the
quenching of magnetism by local quantum spin fluctuations, is
renormalized by the superexchange interaction $J$, so
that $T_K\simeq \left(- c_J J + 2 |\delta t|\right)/k_B $, with $c_J$
a positive numerical constant, $\delta$ the concentration of dopants,
holes $(\delta>0)$ and electrons $(\delta<0)$,  and
$k_B$ the Boltzmann constant. The renormalization term $-c_J J$
depends on $\gamma$. The bandwidth $W^*$ of quasiparticles is about 
$4 k_BT_K$.

When $\gamma \agt W^*$, it follows that $c_J \ll 1$;
the quenching of magnetism by the Kondo effect is weak.  Quasi-two
dimensional thermal spin fluctuations make $T_N$ substantially
reduced; $T_N\simeq 0.2|J|/k_B$ for $\delta=0$, when an
exchange interaction $J_z$ between nearest-neighbor planes is as small
as
$|J_z/J|\simeq 10^{-10}$. This explains observed
$T_N \simeq 300$~K  for $\delta=0$ in cuprates, if we take
$|J|\simeq 0.15$~eV.  Because thermal spin fluctuations vanish at
$T=0$~K, however, an antiferromagnetic state is stabilized  in a wide
range of $\delta$ such as  $0\le |\delta|\alt 0.14$ for $J/|t| =-0.3$. 
When  $T_N\gg T_K$, an antiferromagnetic state is characterized as a
local-moment one. When $T_N\ll T_K$,  it is characterized as an
itinerant-electron one rather than a spin glass. 

When $\gamma\ll W^*$, it follows that $c_J =O(1)$;
the quenching by the Kondo effect is strong.  
When the nesting of the Fermi surface is substantial, the exchange
interaction arising from the virtual exchange of pair excitations of
quasiparticles is also responsible for antiferromagnetic ordering  in
addition to the superexchange interaction. However, an
antiferromagnetic state can only be stabilized for small $|\delta|$
because not only quasi-two dimensional thermal spin fluctuations but
also the Kondo effect with high $T_K$ make $T_N$ substantially reduced
or they destroy antiferromagnetic ordering; the critical $|\delta|$
below which antiferromagnetic ordering appears is smaller for
smaller $\gamma$. An antiferromagnetic state in this case is
characterized as an itinerant-electron one.

The life-time width of quasiparticles arising
from disorder must be a crucial parameter for cuprates. The difference
of disorder must be mainly responsible for the asymmetry of $T_N$
between electron-doped and hole-doped cuprates; disorder must be
relatively larger in electron-doped cuprates than it is in hole-doped
cuprates.  It is interesting to examine if an almost symmetric
behavior of
$T_N$ is restored by preparing hole-doped and electron-doped cuprates
with similar degree of disorder to each other. Because effective
disorder can be enhanced by magnetic fields, it is also interesting to
look for magnetic-field induced antiferromagnetic ordering in cuprates
that exhibit large magnetoresistance.

\begin{acknowledgments}
The author is thankful for useful discussion to K. Kumagai, M. Ido, M. Oda
and N. Momono. This work was supported by a Grant-in-Aid for Scientific
Research (C) Grant No.~13640342 from the Ministry of Education, Cultures,
Sports, Science and Technology of Japan. 
\end{acknowledgments}

\appendix
\section{Scattering potential in disordered Kondo lattices}
\label{SecDisorder}

In disordered Kondo lattices, the mapped Anderson models are
different from site to site. When  the
perturbative expansion for a single-impurity system\cite{Yamada} is
extended to include the site dependence, the selfenergy for the
$j$th site is expanded in such a way that
\begin{eqnarray}
\tilde{\Sigma}_{j\sigma}(\varepsilon \!+\! i0) &=&
\tilde{\Sigma}_{j\sigma}(0) + \left(1-\tilde{\phi}_{j\gamma} \right)\varepsilon
\nonumber \\ &&
-  i \frac{\tilde{\phi}_{js} \!-\! \tilde{\phi}_{j\gamma}}
{2\Delta_j(0)}\left[\varepsilon^2 \!+\! (\pi k_BT)^2 \right] \! +
\cdots ,
\qquad 
\end{eqnarray}
with $\Delta_j(\varepsilon)$ the hybridization energy of the $j$th
Anderson model. 
%%%%%%%%%%%%%%%%%%%%%%%%%%%%%%%%%%%%%%%%%%%%%%%%%%%%%%%%%%%%%%%
When its energy dependence is ignored, 
it follows according to Shiba\cite{shiba} that
\begin{equation}
E_{dj}+\tilde {\Sigma}_{j\sigma}(+i0) = 
\Delta_j(0) \tan \left[
\pi \left( \mbox{$\frac1{2}$} - n_{j\sigma} \right)\right] ,
\end{equation}
with $n_{j\sigma}$ the number of electrons with spin $\sigma$ at the $j$th
site. Here, $E_{dj}$ is the localized-electron level of  
the $j$th mapped Anderson model; it is equal to the band center of the
$t$-$J$ model, so that $E_{dj}=0$. 
%%%%%%%%%%%%%%%%%%%%%%%%%%%%%%%%%%%%%%%%%%%%%%%%%%%%%%%%%%%%%%%%
It follows from Eqs.~(\ref{EqPhiG}) and (\ref{EqPhiS}) that 
$\tilde{\phi}_{j\gamma} \simeq (\pi^2/8)/|1-n_j|$  and 
$\tilde{\phi}_{js} \simeq (\pi^2/4)/|1-n_j|$
for almost half filling 
$n_j \equiv n_{j\uparrow} + n_{j\downarrow} \simeq 1$.
We assume non-magnetic impurities so that
$n_{j\uparrow}=n_{j\downarrow}$.

The average number of $n_j$ and its mean-square deviation are
given by
\begin{eqnarray}
&& \hskip10pt 
n = \int \!\! dx N_{imp}(x) x , 
\\ && \hskip-10pt 
\Delta n^2 = \int \!\! dx N_{imp}(x) \left(x - n
\right)^2 ,
\end{eqnarray}
with  $N_{imp}(n_j)$ the distribution of $n_j$.
We assume that there is no correlation among disorder at 
different sites in our ensemble of disordered systems. 

We can consider the site-dependent part of 
the single-site selfenergy as a static
scattering potential. It is approximately given by
\begin{equation}\label{EqScatt}
V_{j\sigma}(\varepsilon) =
- \left( \frac{\pi \Delta}{2}  
+ \frac{8}{\pi^2}\tilde{\phi}_\gamma^2 \varepsilon \right)
(n_{j} \!-\! n ) 
\frac{1 \!-\! n}{|1 \!-\! n|}+
\cdots .
\end{equation}
Disorder of $n_j$ arises from that of $\Delta_j(\varepsilon)$; in
Eq.~(\ref{EqScatt}),  their site and energy dependences are ignored 
and they are simply dented by $\Delta$. When we treat this
energy-dependent scattering potential in the second-order SSA or the
Born approximation, the coherent part of the  ensemble averaged Green
function is given by
\begin{widetext}
\begin{eqnarray}
\left<g_\sigma^{(0)}(\varepsilon+i0,{\bf k}) \right>_{dis} &=&
 %\frac1{\tilde{\phi}_\gamma} \frac1{\varepsilon+i0  + \mu^* 
 %- \xi({\bf k}) } + \frac1{\tilde{\phi}_\gamma} 
 %\frac1{\varepsilon+i0  + \mu^* - \xi({\bf k})}
 %\Sigma_{\sigma}^{dis} (\varepsilon+i0)  
 %\left<g_\sigma^{(0)}(\varepsilon+i0,{\bf k}) \right>_{dis}
 %\nonumber \\ &=&
 \frac1{\tilde{\phi}_\gamma} \frac1
{\displaystyle \varepsilon+i0  +\mu^* - \xi({\bf k}) - 
(1/\tilde{\phi}_\gamma) \Sigma_{\sigma}^{dis} (\varepsilon+i0)  } ,
\end{eqnarray}
\end{widetext}
with
\begin{equation}
\frac1{\tilde{\phi}_\gamma} \Sigma_{\sigma}^{dis} 
(\varepsilon \!+\! i0) 
= - i \frac{\pi}{\tilde{\phi}_\gamma^2}
\left[ \frac{\pi \Delta}{2} 
\!+\! \frac{8}{\pi^2}\tilde{\phi}_\gamma^2 \varepsilon
\right]^2  \!\!\Delta n^2 \rho_\gamma (\varepsilon) , 
\end{equation}
for $|\varepsilon| \alt k_B T_K $.
Here, 
$\left< \cdots \right>_{dis}$ stands for an ensemble average over
disordered systems.

The bandwidth of quasiparticles is about $4 k_BT_K$, so that typical
life-time width is as large as 
\begin{equation}
\frac1{\tilde{\phi}_\gamma} \mbox{Im}
\Sigma_{\sigma}^{dis} (\pm k_BT_K +i0) 
\simeq - i (64/\pi^3)
\tilde{\phi}_\gamma \Delta n^2  |t| . 
\end{equation}
The energy-independent term can be ignored because 
$\tilde{\phi}_\gamma\gg 1$ for almost half fillings. It is quite
likely that $\tilde{\phi}_\gamma \Delta n^2 =O (1)$ and 
$\gamma/|t|=O(1)$ for almost half fillings. 

%In the presence of nonzero magnetic fields, the distribution of the
%Kondo temperatures causes positive
%magnetoresisitance.\cite{OhkawaPositiveMR}

It is straightforward to extend the above argument to a system in the
presence of magnetic fields\cite{OhkawaPositiveMR} and a system
with magnetic impurities. In such cases,
$n_{i\uparrow}-n_{i\downarrow}$ can be different from site to site; 
$\mbox{Im}\Sigma_{\sigma}^{dis} (\varepsilon+i0)/\tilde{\phi}_\gamma$
can be large even on the chemical potential $(\varepsilon=0)$.

 %  large postive magnetoresistance
 % can  appear.\cite{OhkawaPositiveMR}

%%%%%%%%%%%*%%%%%%%%%%%%%%%%%%%%%%%%%%%%%%%%%%%%%%%%%%
%%%%%%%%%%%%%%%%%%%%%%%%%%%%%%%%%%%%%%%%%%%%%%%%%%%%%
\begin{figure}
\centerline{
\includegraphics[width=7.0cm]{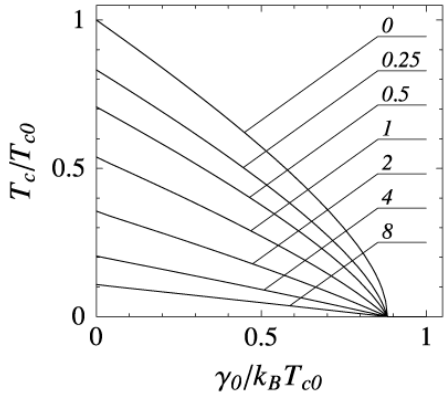}
}
\caption[10]{ 
Superconducting  $T_c$ as a function of
static pair breaking $\gamma_0$ for $\nu=0$, 0.25, 0.5, 1, 2, 4 and 8;
$\gamma=\gamma_0 +\nu k_BT$ is assumed as the total pair breaking.
The result for $\nu=0$ is nothing but that by the AG
theory.\cite{abrikoso-gorkov}
 }
\label{FigTc}
\end{figure} 
%%%%%%%%%%%*%%%%%%%%%%%%%%%%%%%%%%%%%%%%%%%%%%%%%%%%%%
%%%%%%%%%%%%%%%%%%%%%%%%%%%%%%%%%%%%%%%%%%%%%%%%%%%%%

\section{Deviation from the Abrikosov-Gorkov theory}
\label{SecSuper}

The reduction of superconducting $T_c$ by pair breaking is
approximately given by the Abrikosov-Gorkov (AG) 
theory:\cite{abrikoso-gorkov}
\begin{equation}
- \ln \left(\frac{T_c}{T_{c0}} \right) = 
\psi\left(\frac{\gamma}{2\pi k_BT_c} +\frac1{2}\right) 
-\psi\left(\frac1{2}\right) ,
\end{equation}
with $T_{c0}$ critical temperatures in the absence of any pair breaking
or for vanishing life-time widths ($\gamma=0$).   We assume that
life-time widths are given by 
%\begin{equation}
$\gamma= \gamma_0 + \nu k_B T $,
%\end{equation}
where $\gamma_0$ arises from static scatterings by disorder and $\nu
k_B T$ arises from inelastic scatterings by  thermal spin and
superconducting fluctuations; both experimentally and theoretically,
the contribution from thermal fluctuations to $\gamma$ is almost
linear in $T$  except at very low temperatures.  Figure~\ref{FigTc}
shows $T_c$ as a function of $\gamma_0$ for various
$\nu$. It is interesting that the reduction of $T_c$ is almost linear
in $\gamma_0$ for large enough $\nu$ such as $\nu \agt 2$.

According to the previous paper published in 1987,\cite{Ohkawa87SC-2}
the ratio of
$\epsilon_G(0)/k_BT_c$, with $\epsilon_G(0)$ superconducting gaps at
$T=0$~K, is as large as $4.35$ for
$d\gamma$ wave. However, observed ratios are larger than that, for
example,
$\epsilon_G(0)/k_BT_c \simeq 8$ for optimal-doped cuprates and 
$\epsilon_G(0)/k_BT_c \gg 8$ for under-doped cuprates. This discrepancy
can be explained in terms of the temperature dependent pair breaking,
because it reduces $T_c$ but it does not reduce
$\epsilon_G(0)$.\cite{OhkawaPseudogap}  Considering
observed ratios of $\epsilon_G(0)/k_BT_c$ and Fig.~\ref{FigTc}, we
argue that
$\nu=1\mbox{--} 2$ for optimal-doped cuprates and $\nu\gg 2$ for
under-doped cuprates. When
$\nu\agt 2$, the reduction of $T_c$ is almost linear in  $\gamma_0$.
According to a recent observation,\cite{alloul} in actual, the
reduction of $T_c$ is almost linear in the dose of electron
irradiation.


\begin{thebibliography}{}
\bibitem{comSSA}
Such an SSA may
include the renormalization of single-site terms by intersite
effects, as is shown in Sec.~\ref{SecRenomalizedSSA} of this paper.
\bibitem{ComGamma} 
Because of this simplification,
we can carry out analytically  summations over
Matsubara's frequencies in many equations of this paper.
\bibitem{CommentSC}
Two mechanisms of the formation of Cooper pairs by the
superexchange interaction and the spin-fluctuation mediated
interaction are essentially the same as each other. When low-energy
spin fluctuations, whose energies are much smaller than $U$, are only
considered, however, they are physically different from
each other.
\bibitem{comment-Ws}
In Eqs.~(\ref{EqTK2}) and (\ref{EqTKdef}), we should use a
multiplication factor slightly different from  $r$ because
$\tilde{\phi}_s$ appearing there is a purely single-site property.  In
the first term of Eq.~(\ref{EqReplace}), we should also use another
one slightly different from $r$  because no $\tilde{\phi}_s$ appears
there. However, we  assume  the same $r$ in every equation for the
sake of simplicity.
\bibitem{ComGammaE-dep}  
When the energy dependence of $\gamma$ is taken into account $k_F l$
can be larger than the estimation of this paper is.
\end{thebibliography}
\end{document}